\documentclass[12pt]{article}

\usepackage{amssymb}
\usepackage{amsmath}
\usepackage{color}
\usepackage{subcaption}
\usepackage{framed}
\usepackage{mdframed}
\usepackage{adjustbox}
\usepackage{graphicx}
\usepackage{float}
\usepackage{comment}
\usepackage{fancyhdr}

    \newtheorem{theorem}{Theorem}[section]

\newcommand{\LM}[1]{\hbox{\vrule width.2pt \vbox to#1pt{\vfill \hrule width#1pt height.2pt}}}
\newcommand{\LL}{{\mathchoice
{\,\LM7\,}{\,\LM7\,}{\,\LM5\,}{\,\LM{3.35}\,}}}

\graphicspath {
    {./}
    {./}
}

\title{Data-Driven Computing with Noisy Material-Data Sets}

\author{
        T.~Kirchdoerfer and M.~Ortiz\\
        Graduate Aeronautical Laboratories \\
        California Institute of Technology \\
        Pasadena, CA 91125, USA}

\begin{document}

\maketitle

\begin{abstract}
We formulate a Data Driven Computing paradigm, termed max-ent Data Driven Computing, that generalizes distance-minimizing Data Driven Computing and is robust with respect to outliers. Robustness is achieved by means of clustering analysis. Specifically, we assign data points a variable relevance depending on distance to the solution and on maximum-entropy estimation. The resulting scheme consists of the minimization of a suitably-defined free energy over phase space subject to compatibility and equilibrium constraints. Distance-minimizing Data Driven schemes are recovered in the limit of zero temperature. We present selected numerical tests that establish the convergence properties of the max-ent Data Driven solvers and solutions.
\end{abstract}


\section{Introduction}

Despite the phenomenal growth of scientific computing over the past 50 years, several stubborn challenges have remained foci of extensive research to this day. One of those challenges is {\sl material modelling}. The prevailing and classical scientific computing paradigm has been to calibrate empirical material models using observational data and then use the calibrated material models in calculations. This process of modelling inevitably adds error and uncertainty to the solutions, especially in systems with high-dimensional phase spaces and complex material behavior. This modelling error and uncertainty arises mainly from imperfect knowledge of the functional form of the material laws, the phase space in which they are defined, and from scatter and noise in the experimental data. Simultaneously, advances in experimental science over the past few decades have changed radically the nature of science and engineering from {\sl data-starved} fields to, increasingly, {\sl data-rich} fields, thus opening the way for the application of the emerging field of {\sl Data Science} to science and engineering. Data Science currently influences primarily non-STEM fields such as marketing, advertising, finance, social sciences, security, policy, and medical informatics, among others. By contrast, the full potential of Data Science as it relates to science and engineering has yet to be explored and realized.

The present work is concerned with the development of a Data Science paradigm, to be referred to as {\sl Data Driven Computing}, tailored to scientific computing and analysis, cf.~\cite{RN48}. Data Driven Computing aims to formulate initial-boundary-value problems, and corresponding calculations thereof, directly from material data, thus bypassing the empirical material modelling step of traditional science and engineering altogether. In this manner, material modelling empiricism, error and uncertainty are eliminated entirely and no loss of experimental information is incurred. Here, we extend earlier work on Data Driven Computing \cite{RN48} to random material data sets with finite probability of {\sl outliers}. We recall that the Data Driven Computing paradigm formulated in \cite{RN48}, or {\sl distance-minimizing} Data Driven Computing, consists of identifying as the best possible solution the point in the material data set that is closest to satisfying the field equations of the problem. Equivalently, the distance-minimizing Data Driven solution can be identified with the point in phase space that satisfies the field equations and is closest to the material data set. It can be shown \cite{RN48} that distance-minimizing Data Driven solutions converge with respect to uniform convergence of the material set. However, distance-minimizing Data Driven solutions can be dominated by {\sl outliers} in cases in which the material data set does not converge uniformly. Distance-minimizing Data Driven solvers are sensitive to outliers because they accord overwhelming influence to the point in the material data set that is closest to satisfying the field equations, regardless of any clustering of the material data points.

The central objective of the present work is to develop a new Data Driven Computing paradigm, to be called max-ent Data Driven Computing, that generalizes distance-minimizing Data Driven Computing and is robust with respect to outliers. Robustness is achieved by means of clustering analysis. Specifically, we assign data points a variable relevance depending on distance to the solution and through maximum-entropy estimation. The resulting scheme consists of the minimization of a free energy over phase space subject to compatibility and equilibrium constraints. We note that this problem is of non-standard type, in that the relevant free energy is a function of state defined over phase space, i.~e., a joint function of the driving forces and fluxes of the system. Max-ent Data Driven solutions are robust with respect to outliers because a cluster of data points can override an outlying data point even if the latter is closer to the constraint set that any point in the cluster. The distance-minimizing Data Driven schemes \cite{RN48} are recovered in the limit of zero temperature. We also develop a simulated annealing scheme that, through an appropriate annealing schedule zeros in on the most relevant data cluster and the attendant solution. We assess the convergence properties of max-ent Data Driven solutions and simulated annealing solver by means of numerical testing.

The paper is organized as follows. In Section~\ref{cro6Fo}, we begin by laying out the connection between Data Science and Scientific Computing that provides the conceptual basis for Data Driven Computing. In Section~\ref{P4oupR}, we turn attention to random material data sets that may contain outliers, or points far removed from the general clustering of the material data points, with finite probability and develop max-ent Data Driven solvers by an appeal to Information Theory and maximum-entropy estimation. In Section~\ref{h9eslE}, we develop a simulated annealing solver that zeros in on the solution, which minimizes a suitably-defined free energy over phase space by progressive quenching. In Section~\ref{p6oeWo}, we present numerical tests that assess the convergence properties of max-ent Data Driven solutions with respect to uniform convergence of the material data set. We also demonstrate the performance of Data Driven Computing when the material behavior itself is random, i.~e., defined by a probability density over phase space. Finally, concluding remarks and opportunities for further development of the Data Driven paradigm are presented in Section~\ref{1iaFRl}.

\section{The Data Driven Science paradigm}\label{cro6Fo}

In order to understand the hooks by which Data Science may attach itself to Scientific Computing, it helps to review the structure of a typical scientific calculation. Of special import to the present discussion is the fundamentally different roles that conservation and material laws play in defining that structure, with the former setting forth hard universal or material-independent constraints on the states attainable by the system and the latter bringing in material specificity open to empirical determination and sampling.

\subsection{The 'anatomy' of boundary-value problems}

We begin by noting that the field theories that provide the basis for scientific computing have a common general structure. Perhaps the simplest field theory is potential theory, which arises in the context of Newtonian mechanics, hydrodynamics, electrostatics, diffusion, and other fields of application. In this case, the field $\varphi$ that describes the global state of the system is scalar. The {\sl localization law} that extracts from $\varphi$ the {\sl local state} at a given material point is  $\epsilon = \nabla\varphi$, i.~e., the localization operator is simply the gradient of the field, together with essential boundary conditions of the Dirichlet type. The corresponding conjugate variable is the {\sl flux} $\sigma$. The flux satisfies the {\sl conservation equation} $\nabla \cdot \sigma = \rho$, where $\nabla\cdot$ is the divergence operator and $\rho$ is a source density, together with natural boundary conditions of the Neumann type. The pair $z = (\epsilon, \sigma)$  describes the local state of the system at a given material point and takes values in the product space $Z = \mathbb{R}^n \times \mathbb{R}^n$, or {\sl phase space}. We note that the phase space, localization and conservation laws are universal, i.~e., material independent. We may thus define a material-independent {\sl constraint set} $C$ to be the set of local states $z = (\epsilon, \sigma)$ consistent with the localization and conservation laws, including corresponding essential and natural boundary conditions.

The localization and conservation laws are closed by appending an appropriate material law. In general, material laws express a relation between fluxes and corresponding {\sl driving forces}. In field theories, the assumption is that local states supply the forces driving the fluxes, leading to material laws of the form $\sigma(\epsilon)$. Often, such material laws are only known imperfectly through a material data set $E$ in phase space $Z$ that collects the totality of our empirical knowledge of the material. Thus, suppose that, in contrast to the classical formulation of initial-boundary-value problems in science and engineering, the material law is imperfectly characterized by a material data point set $E$. A typical material data set then consists of a finite number of local states, $E=((\epsilon_i,\sigma_i ),\ i=1,\dots,N)$. Evidently, for a material data set of this type, the intersection $E\cap C$ is likely to be empty, i.~e., there may be no points in the material data set that are compatible with the localization and conservation laws, even in cases when solutions could reasonably be expected to exist. It is, therefore, necessary to replace the overly-rigid characterization of the solution set $S = E \cap C$ by a suitable relaxation thereof.

\subsection{Distance-minimizing Data Driven schemes}

One such relaxed formulation of Data Driven Computing \cite{RN48} consists of accepting as the best possible solution the point $z_i=(\epsilon_i, \sigma_i)$ in the material data set $E$ that is closest to the constrained set $C$, i.~e., the point that is closest to satisfying the localization and conservation laws. Closeness is understood in terms of some appropriate distance $d$ defined in phase space $Z$. The corresponding distance from a local state $z$ to the material data set $E$ is, then: $d(z,E)=\min_{y\in E} d(z,y)$,  and the optimal solution is the solution of the minimum problem: $\min_{z\in C} d(z,E)$. Evidently, the data driven problem can also be directly formulated as the double minimization problem: $\min_{z\in C} \min_{y\in E} d(z,y)$. Inverting the order of minimization, we obtain the equivalent data driven problem: $\min_{y\in E} \min_{z\in C} d(z,y)$, or: $\min_{y\in E} d(y,C)$. This reformulation identifies the data driven solution as the point $y$ in the constraint set $C$ that is closest to the material data set $E$.

\begin{figure}[h]
\begin{center}
    \includegraphics[width=0.75\textwidth]{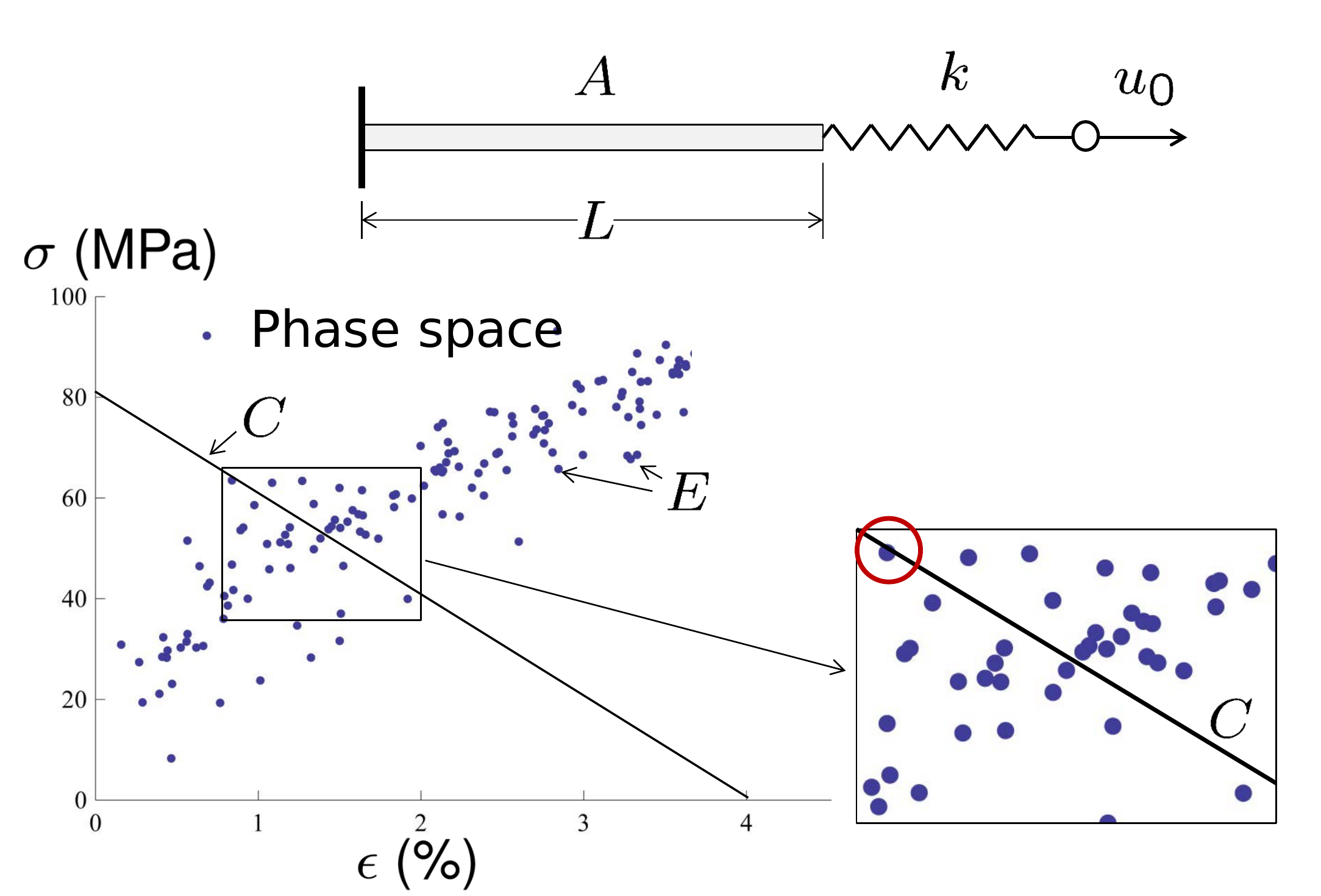}
    \caption{Bar loaded by soft device. The data driven solution is the point in the material data set (circled in red) that is closest to the constraint set.}
    \label{DQ3jMK}
\end{center}
\end{figure}

\subsection{An elementary example}

The distance-minimizing Data Driven Computing paradigm just outlined is illustrated in Fig.~\ref{DQ3jMK} by means of the elementary example of a elastic bar deforming uniformly under the action of a well-calibrated loading device. In this example, phase space $Z$ is the $(\epsilon,\sigma)$-plane, the material data set is a point set $E$ in phase space and the constraint set is a straight line $C$ of slope and location determined by the stiffness $k$ of the loading device and the applied displacement $u_0$. In general, the constraint set $C$ and the material data set $E$ may have empty intersection. However, the distance-minimizing Data Driven solution is well defined, though not necessarily uniquely, as the point of the material data set that is closest to the constraint set, circled in red in Fig.~\ref{DQ3jMK}. The fundamental property of the distance-minimizing Data Driven scheme thus defined is that it makes direct use of the material data set in calculations, entirely bypassing the intermediate modelling step of conventional material identification. In the simple example of the bar loaded by a soft device, it is clear that the distance-minimizing Data Driven solution converges to a classical solution if the data traces a graph in phase space with increasing sampling density, which is a {\sl sanity-check} requirement.

\subsection{Uniform convergence}

The variational structure of distance-min\-imiz\-ing Data Driven problems confers additional robustness to the solvers and renders them amenable to analysis. By exploiting this connection, it can be shown \cite{RN48} that distance-minimizing Data Driven solutions converge to classical solutions when the data set approximates a limiting graph in phase space with increasing fidelity, a test case that provides a {\sl sanity check}. Specifically, suppose that the limiting material law is represented by a graph $E$ in phase space, and that a sequence $(E_k )$ of material data sets is such that: i) there is a sequence $\rho_k \downarrow 0$ such that $ {\rm dist}(z, E_k) \leq \rho_k$, for all $z \in E$, and ii) there is a sequence $t_k \downarrow 0$ such that ${\rm dist}(z_k, E) \leq t_k$, for all $z_k \in E_k$. Then, with an additional transversality assumption between the data and the constraint set, it follows \cite{RN48} that the corresponding sequence $(z_k)$ of distance-minimizing Data Driven solutions converges to the solution $z$ of the classical problem defined by the classical material law $E$. In addition, if the discrete problem is the result of spatial discretization, e.~g., by means of the finite element method, then, under the same assumptions, the sequence $(u_k)$ of solutions corresponding to a sequence $(E_k)$ of material data sets converges in norm to the classical solution $u$ of the boundary value problem defined by the classical material law $E$.

\section{Probabilistic Data Driven schemes}
\label{P4oupR}

In practice, material data sets may be random by virtue of inherent stochasticity of the material behavior, experimental scatter inherent to the method of measurement, specimen variability and other factors. Under these conditions, the material data set may contain outliers, or points far removed from the general clustering of the material data points, with finite probability. If one of these outliers happens to be close to the constraint set, it may unduly dominate the distance-minimizing Data Driven solution described in the foregoing. Thus, such solvers, while well-behaved in applications with material data sets with uniformly bounded scatter, may not be sufficiently robust with respect to persistent outliers in other applications. This limitation of distance-minimizing Data Driven solvers points to the need to investigate problems with random material data sets from a probabilistic perspective, with a view to ranking the data points by relevance and importance and understanding the probability distribution of outcomes of interest.

\subsection{Data clustering}

Distance-minimizing Data Driven solvers are sensitive to outliers because, for any given test solution $z$ in phase space, they accord overwhelming influence to the nearest point in the material data set, regardless of any clustering of the points. Cluster analysis provides a means of mitigating the influence of individual material data points and building notions of data clustering into the Data Driven solver.  Cluster analysis can be based on fundamental concepts of Information Theory such as maximum-entropy estimation \cite{RN129}. Specifically, we wish to quantify how well a point $z$ in phase space is represented by a point $z_i$ in a material data set $E=(z_1,\dots z_n)$. Equivalently, we wish to quantify the relevance of a point $z_i$ in the material data set to a given point $z$ in phase space. We measure the relevance of points $z_i$ in the material data set by means of weights $p_i\in [0,1]$ with the property
\begin{equation}
    \sum_{i=1}^n p_i=1 .
\end{equation}
We wish the ranking by relevance of the material data points to be unbiased. It is known from Information Theory that the most unbiased distribution of weights is that which maximizes Shannon's information entropy \cite{RN132,RN133,RN134}
\begin{equation}\label{sIa8wo}
    H(p)
    =
    - \sum_{i=1}^n p_i \log p_i
\end{equation}
with the extension by continuity: $0 \log 0 = 0$. In addition, we wish to accord points distant from $z$ less weight than nearby points, i.~e., we wish the cost function
\begin{equation}\label{chOa6r}
    U(z,p)
    =
    \sum_{i=1}^n p_i d^2(z-z_i)
\end{equation}
to be as small as possible. These competing objectives can be combined in the sense of Pareto optimality. The Pareto optima are solutions of the problem
\begin{eqnarray}\label{woa4oA}
    \mbox{For fixed $z$, minimize:} \quad
    && \beta U(z,p) - H(p) \\
    \mbox{subject to:} \quad && p_i \ge 0, \ i=1,\dots,n; \quad \sum_{i=1}^N p_i = 1 ,
\end{eqnarray}
where $\beta \in (0,+\infty)$ is a Pareto weight. The solution to this problem is given by the Bolzmann distribution
\begin{subequations}\label{tR2abr}
\begin{align}
    &
    p_i(z,\beta) = \frac{1}{Z(z)} {\rm e}^{-(\beta/2) d^2 (z,z_i)} ,
    \\ & \label{kTKH4K}
    Z(z,\beta) = \sum_{i=1}^n {\rm e}^{-(\beta/2) d^2 (z,z_i)} .
\end{align}
\end{subequations}
The corresponding max-ent Data Driven solver now consists of minimizing the free energy
\begin{equation}\label{tpG2dU}
    F(z,\beta) = - \frac{1}{\beta} \log Z(z,\beta) ,
\end{equation}
over the constraint set $C$, i.~e.,
\begin{equation}\label{SoesI5}
    z \in {\rm argmin} \{ F(z',\beta),\ z' \in C \} .
\end{equation}

We note that ${\beta^{-1/2}}$ represents the width of the Bolzmann distribution (\ref{tR2abr}) in phase space. Thus, points in the data set at a distance to $z$ large compared to ${\beta^{-1/2}}$ have negligible influence over the solution. Conversely, the solution $z$ is dominated by the local cluster of data points in the ${\beta^{-1/2}}$-neighborhood of $z$. In particular, outliers, or points outside that neighborhood, have negligible influence over the solution.

For a compact material point set $E$ in a finite-dimensional phase space, the existence of solutions of problem (\ref{SoesI5}) is ensured by the Weierstrass extreme value theorem. We also note that the distance-minimizing Data Driven scheme \cite{RN48} is recovered in the limit of $\beta \to +\infty$. By analogy to statistical thermodynamics, max-ent Data Driven Computing may be regarded as a {\sl thermalized} extension of distance-minimizing Data Driven Computing. For finite $\beta$, all points in the material data set influence the solution, but their corresponding weights diminish with distance to the solution. In particular, the addition of an outlier that is marginally closer to the constraint set $C$ than a large cluster of material data points does not significantly alter the solution, as desired.

\section{Numerical implementation}
\label{h9eslE}

We recall that the max-ent Data Driven problem of interest is to minimize the free energy $F(z)$ (\ref{tpG2dU}) over the constraint set $C$. The corresponding optimality condition is
\begin{equation}\label{NnS2s3}
    \frac{\partial F}{\partial z}(z,\beta)
    \perp
    C ,
\end{equation}
where $\perp$ denotes orthogonality. Assuming
\begin{equation}\label{4IEbri}
    d(z,z') = | z - z' | ,
\end{equation}
with $| \cdot |$ the standard norm in $\mathbb{R}^n$, we compute
\begin{equation}
    \frac{\partial F}{\partial z}(z,\beta)
    =
    \sum_{i=1}^n
    p_i(z,\beta)
    (z - z_i)
    =
    z
    -
    \sum_{i=1}^n
    p_i(z,\beta)
    z_i .
\end{equation}
Inserting this identity into (\ref{NnS2s3}), we obtain
\begin{equation}
    z
    -
    \sum_{i=1}^n
    p_i(z,\beta)
    z_i
    \perp
    C ,
\end{equation}
which holds if and only if
\begin{equation}\label{AVsY3J}
    z
    =
    P_C\left(
        \sum_{i=1}^n
        p_i(z,\beta)
        z_i
    \right) ,
\end{equation}
where $P_C$ is the closest-point projection to $C$.
For instance, if $C = \{ f(z) = 0 \}$ for some constraint function $f(z)$, (\ref{NnS2s3}) may be expressed as
\begin{subequations}\label{uy55LT}
\begin{align}
    &
    \frac{\partial F}{\partial z}(z,\beta)
    =
    \lambda \frac{\partial f}{\partial z}(z) ,
    \\ &
    f(z) = 0 ,
\end{align}
\end{subequations}
where $\lambda$ is a Lagrange multiplier.

The essential difficulty inherent to problem (\ref{NnS2s3}), or (\ref{uy55LT}), is that, in general, the free energy function $F(\cdot,\beta)$ is strongly non-convex, possessing multiple wells centered at the data points in the material data set. Under these conditions, iterative solvers may fail to converge or may return a local minimizer, instead of the global minimizer of interest.

We overcome these difficulties by recourse to {\sl simulated annealing} \cite{RN124}. The key observation is that the free energy $F(\cdot,\beta)$ is convex for sufficiently small $\beta$. Indeed, a straightforward calculation using (\ref{4IEbri}) gives the Hessian matrix as
\begin{equation}
\begin{split}
    \frac{\partial^2 F}{\partial z\partial z}(z)
    & =
    I
    -
    \beta
    \sum_{i=1}^n
    p_i(z)
    (z - z_i)\otimes (z - z_i)
    \\ & +
    \beta
    \left(
        \sum_{i=1}^n p_i(z) (z - z_i)
    \right)
    \otimes
    \left(
        \sum_{j=1}^n p_j(z) (z - z_j)
    \right)
\end{split}
\end{equation}
Evidently, in the limit of $\beta \to 0$ the Hessian reduces to the identity and the free energy is convex. Indeed, it follows from (\ref{kTKH4K}) that, for $\beta\to 0$,
\begin{equation}
    F(z,\beta)
    -
    \frac{1}{\beta}
    \log\frac{1}{n}
    \sim
    \frac{1}{n} \sum_{i=1}^n\frac{1}{2} d^2 (z,z_i) .
\end{equation}
The main idea behind simulated annealing is, therefore, to initially set $\beta$ sufficiently small that $F(\cdot,\beta)$ is convex and subsequently increase it according to some appropriate annealing schedule, with a view to guiding the solver towards the absolute minimizer.

\subsection{Fixed-point iteration}

We begin by noting that eq.~(\ref{AVsY3J}) conveniently defines the following fixed-point iteration,
\begin{equation}\label{uM3Vd6}
    z^{(k+1)}
    =
    P_C\left(
        \sum_{i=1}^n
        p_i(z^{(k)},\beta)
        z_i
    \right) .
\end{equation}
We recall that fixed-point iterations $z \leftarrow f(z)$ converge if the mapping $f(z)$ is contractive. Since $P_C$ is an orthogonal projection, it is contractive if the constraint set $C$ is convex, which we assume henceforth. Under this assumption, the mapping (\ref{uM3Vd6}) is contractive if and only if the mapping
\begin{equation}\label{a6Hb7W}
    z
    \mapsto
    \sum_{i=1}^n p_i(z,\beta) z_i
    \equiv
    g(z,\beta)
    =
    z - \frac{\partial F}{\partial z}(z,\beta)
\end{equation}
is contractive. Conditions ensuring the contractivity of $g(\cdot,\beta)$ are given by the following theorem.

\begin{theorem}\label{h6H7DJ}
Suppose that
\begin{equation}\label{PmCj4W}
    \frac{1}{\beta}
    <
    \sum_{i=1}^n
    p_i(z,\beta)
    | z_i-\bar{z} |^2 ,
\end{equation}
where
\begin{equation}\label{hxT72w}
    \bar{z} = \sum_{i=1}^n p_i(z,\beta) z_i .
\end{equation}
Then, $g(\cdot,\beta)$ is contractive in a neighborhood of $z$.
\end{theorem}

{\bf Proof.} From the definition (\ref{a6Hb7W}) of $g(z,\beta)$, we have, after a trite calculation,
\begin{equation}
    \frac{\partial g}{\partial z}(z)
    =
    \beta
    \sum_{i=1}^n
    p_i(z,\beta)
    (z_i-\bar{z}) \otimes (z_i-\bar{z}) .
\end{equation}
Let $u \in Z$, $|u|=1$. Then,
\begin{equation}\label{5EC5zy}
    u^T \frac{\partial g}{\partial z}(z) u
    =
    \beta
    \sum_{i=1}^n
    p_i(z,\beta)
    \big((z_i-\bar{z}) \cdot u\big)^2
    \leq
    \beta
    \sum_{i=1}^n
    p_i(z,\beta)
    | z_i-\bar{z} |^2 .
\end{equation}
Therefore, by the implicit function theorem, contractivity in a neighborhood of $z$ follows if
\begin{equation}
    \beta
    \sum_{i=1}^n
    p_i(z,\beta)
    | z_i-\bar{z} |^2
    <
    1 ,
\end{equation}
or, equivalently, if (\ref{PmCj4W}) holds. \hfill$\square$

\subsection{Simulated annealing}
\label{9iechI}

The general idea of simulated annealing is to evolve the reciprocal temperature jointly with the fixed point iteration according to an appropriate annealing schedule, i.~e., we modify (\ref{uM3Vd6}) to
\begin{equation}
    z^{(k+1)}
    =
    P_C\left(
        \sum_{i=1}^n
        p_i(z^{(k)},\beta^{(k)})
        z_i
    \right) .
\end{equation}
An effective annealing schedule is obtained by selecting $\beta^{(k+1)}$ so as to ensure local contractivity of the fixed-point mapping. An appeal to Theorem~\ref{h6H7DJ} suggests the schedule
\begin{equation}\label{Vo9Xoa}
    \frac{1}{\beta^{(k+1)}}
    =
    \sum_{i=1}^n
    p_i(z^{(k)},\beta^{(k)})
    | z_i-\bar{z}^{(k)} |^2 ,
\end{equation}
with the initial reciprocal temperature $\beta^{(0)}$ chosen small enough that the mapping $g(\cdot, \beta^{(0)})$ is contractive everywhere. An estimate that provides a suitable initial reciprocal temperature $\beta^{(0)}$ is provided by the following theorem.

\begin{theorem}\label{zEN8h9}
Suppose that
\begin{equation}\label{8ugL3r}
    \frac{1}{\beta}
    >
    \frac{1}{n}
    \sum_{i=1}^n |z - z_i|^2 ,
\end{equation}
for all $z \in \Omega \subset Z$. Then, $F(\cdot,\beta)$ is convex in $\Omega$.
\end{theorem}

{\bf Proof}. Fix $z \in \Omega$ and $\beta > 0$ and let $u \in Z$ be an arbitrary unit vector in phase space.
We have
\begin{equation}
    u^T \frac{\partial^2 F}{\partial z\partial z} u
    =
    1
    -
    \beta
    \sum_{i=1}^n
    p_i(z,\beta)
    \big((z - z_i)\cdot u\big)^2
    +
    \beta
    \left(
        \sum_{i=1}^n p_i(z,\beta) (z - z_i)\cdot u
    \right)^2 ,
\end{equation}
which gives the lower bound
\begin{equation}
    u^T \frac{\partial^2 F}{\partial z\partial z} u
    \geq
    1
    -
    \beta
    \sum_{i=1}^n
    p_i(z,\beta)
    \big((z - z_i)\cdot u\big)^2 .
\end{equation}
Maximizing the bound with respect to $u$, we have
\begin{equation}
    \sum_{i=1}^n
    p_i(z,\beta)
    \big((z - z_i) \cdot u\big)^2
    \leq
    \sum_{i=1}^n
    p_i(z,\beta)
    |z - z_i|^2 ,
\end{equation}
and, hence,
\begin{equation}\label{J6tqQD}
    u^T \frac{\partial^2 F}{\partial z\partial z} u
    \geq
    1
    -
    \beta
    \sum_{i=1}^n p_i(z,\beta) |z - z_i|^2 .
\end{equation}
From the max-ent optimality of $p_i(z,\beta)$, we additionally have
\begin{equation}
    \frac{\beta}{2}
    \sum_{i=1}^n p_i(z,\beta) |z - z_i|^2 + \sum_{i=1}^n p_i(z,\beta) \log p_i(z,\beta)
    \leq
    \frac{\beta}{2}
    \sum_{i=1}^n p'_i |z - z_i|^2 + \sum_{i=1}^n p'_i \log p'_i ,
\end{equation}
for all $(p'_i)$ such that
\begin{equation}
    p'_i \geq 0 ,
    \quad
    \sum_{i=1}^n  p'_i = 1 .
\end{equation}
Testing with $p'_i = 1/n$, we obtain
\begin{equation}\label{BFn8KK}
    \frac{\beta}{2}
    \sum_{i=1}^n p_i(z,\beta) |z - z_i|^2
    +
    \sum_{i=1}^n p_i(z,\beta) \log p_i(z,\beta)
    \leq
    \frac{\beta}{2}
    \frac{1}{n}
    \sum_{i=1}^n |z - z_i|^2
    +
    \log \frac{1}{n}
\end{equation}
But, by Jensen's inequality,
\begin{equation}
    \log \frac{1}{n}
    \leq
    \sum_{i=1}^n p_i(z,\beta) \log p_i(z,\beta) ,
\end{equation}
which, in conjunction with (\ref{BFn8KK}) gives
\begin{equation}
    \sum_{i=1}^n p_i(z,\beta) |z - z_i|^2
    \leq
    \frac{1}{n}
    \sum_{i=1}^n |z - z_i|^2 .
\end{equation}
Inserting this estimate in (\ref{J6tqQD}) gives
\begin{equation}
    u^T \frac{\partial^2 F}{\partial z\partial z} u
    \geq
    1
    -
    \frac{\beta}{n}
    \sum_{i=1}^n |z - z_i|^2 .
\end{equation}
From this inequality we conclude that
\begin{equation}
    u^T \frac{\partial^2 F}{\partial z\partial z} u
    \geq
    0
\end{equation}
for all unit vectors $u$ in phase space if
\begin{equation}
    1
    -
    \frac{\beta}{n}
    \sum_{i=1}^n |z - z_i|^2
    \geq
    0 ,
\end{equation}
or, equivalently, if inequality (\ref{8ugL3r}) is satisfied. \hfill$\square$

One way to use the preceding theorem is as follows. Suppose that $\Omega \subset Z$ contains the data set and the expected solution. Then, for any $z \in \Omega$, we have
\begin{equation}
    |z - z_i| \leq {\rm diam}(\Omega) ,
\end{equation}
where ${\rm diam}(\Omega)$ is the diameter of $\Omega$. Additionally, we have that
\begin{equation}
    \frac{1}{n}
    \sum_{i=1}^n |z - z_i|^2
    \geq
    \frac{1}{n} {\rm diam}^2(\Omega) ,
    \label{DiamEst}
\end{equation}
and the right-hand side of this inequality supplies a conservative estimate of the $\beta$ threshold for convexity.

As already noted, the max-ent Data Driven solution is controlled by its local ${\beta^{-1/2}}$-neighborhood of points in the data set. Thus, initially the annealing schedule casts a broad net and all points in the data set are allowed to influence the solution. As $\beta$ grows, that influence is restricted to an increasingly smaller cluster of data points. For large $\beta$, the solution is controlled by the points in a certain local neighborhood of the data set determined by the annealing iteration. In particular, the influence of outliers in the data set is eliminated.

\section{Numerical tests}\label{p6oeWo}

We test the properties of max-ent Data Driven Computing by means of the simple example of truss structures. Trusses are assemblies of articulated bars that deform in uniaxial tension or compression. Thus, conveniently, in a truss the material behavior of a bar $e$ is characterized by a simple relation between the uniaxial strain $\varepsilon_e$ and uniaxial stress $\sigma_e$ in the bar. We refer to the space of pairs $z_e = (\varepsilon_e, \sigma_e)$ as the {\sl phase space} of bar $e$. We assume that the behavior of the material of each bar $e=1,\dots,m$, where $m$ is the number of bars in the truss, is characterized by---possibly different---local data sets $E_e$ of pairs $z_e$, or {\sl local states}. For instance, each point in the data set may correspond, e.~g., to an experimental measurement. A typical data set is notionally depicted in Fig.~\ref{DQ3jMK}. The global data set is then the cartesian product $E = \prod_{e=1}^m E_e$ of all local data sets.

The state $z = (z_e)_{e=1}^m$ of the truss is subject to the compatibility and equilibrium constraints
\begin{subequations}
\begin{align}
    & \label{Bd2bYR}
    \epsilon_e = B_e u ,
    \\ & \label{j8dCjM}
    \sum_{e=1}^m B_e^T w_e \sigma_e = f  ,
\end{align}
\end{subequations}
where $u$ is the array of nodal displacements, $f$ is the array of applied nodal forces, the matrices $(B_e)_{e=1}^m$ encode the geometry and connectivity of the truss members and $w_e$ is the volume of member $e$.

We may metrize the local phase spaces of each member of the truss by means of Euclidean distances derived from the norms
\begin{equation}\label{fou9Oe}
    | z_e |_e
    =
    \left(
        \mathbb{C} \epsilon_e^2
        +
        \mathbb{C}^{-1} \sigma_e^2
    \right)^{1/2} ,
\end{equation}
for some positive constant $\mathbb{C}$. We may then metrize the global state of the truss by means of the global norm
\begin{equation}\label{9oakLa}
    | z |
    =
    \Big( \sum_{e=1}^m w_e | z_e |_e^2 \Big)^{1/2}
    =
    \left(
        \sum_{e=1}^m
        w_e
        \Big(
            \mathbb{C} \epsilon_e^2
            +
            \mathbb{C}^{-1} \sigma_e^2
        \big)
    \right)^{1/2}
\end{equation}
and the associated distance (\ref{4IEbri}). For a truss structure, the point in $C$ closest to a given point $z^*$ in phase space follows from the stationarity condition
\begin{equation}
    \delta
    \left\{
        \sum_{e=1}^m
        w_e
        \left(
            \frac{\mathbb{C}}{2} (B_e u - \epsilon_e^*)^2
            +
            \frac{\mathbb{C}^{-1}}{2} (\sigma_e - \sigma_e^*)^2
        \right)
        +
        \Big( f - \sum_{e=1}^m w_e B_e^T \sigma_e \Big)^T \lambda
    \right\}
    =
    0 ,
\end{equation}
where $\lambda$ is an array of Lagrange multiplier enforcing the equilibrium constraints. The corresponding Euler-Lagrange equations are
\begin{subequations}
\begin{align}
    &
    \sum_{e=1}^m w_e B_e^T \mathbb{C} (B_e u - \epsilon_e^*) = 0 ,
    \\ &
    \mathbb{C}^{-1} (\sigma_e - \sigma_e^*) - B_e \lambda = 0 ,
    \\ &
    \sum_{e=1}^m w_e B_e^T \sigma_e = f ,
\end{align}
\end{subequations}
or
\begin{subequations}
\begin{align}
    &
    \left(\sum_{e=1}^m w_e B_e^T \mathbb{C} B_e \right) u
    =
    \sum_{e=1}^m w_e B_e^T \mathbb{C} \epsilon_e^* ,
    \\ &
    \left(\sum_{e=1}^m w_e B_e^T \mathbb{C} B_e \right) \lambda
    =
    f - \sum_{e=1}^m w_e B_e^T\sigma_e^* ,
\end{align}
\end{subequations}
which define two standard truss equilibrium problems for the linear reference material of modulus $\mathbb{C}$.

\begin{figure}[h]
\begin{subfigure}[]{0.43\textwidth}
    \centering
    \includegraphics[width=\textwidth]{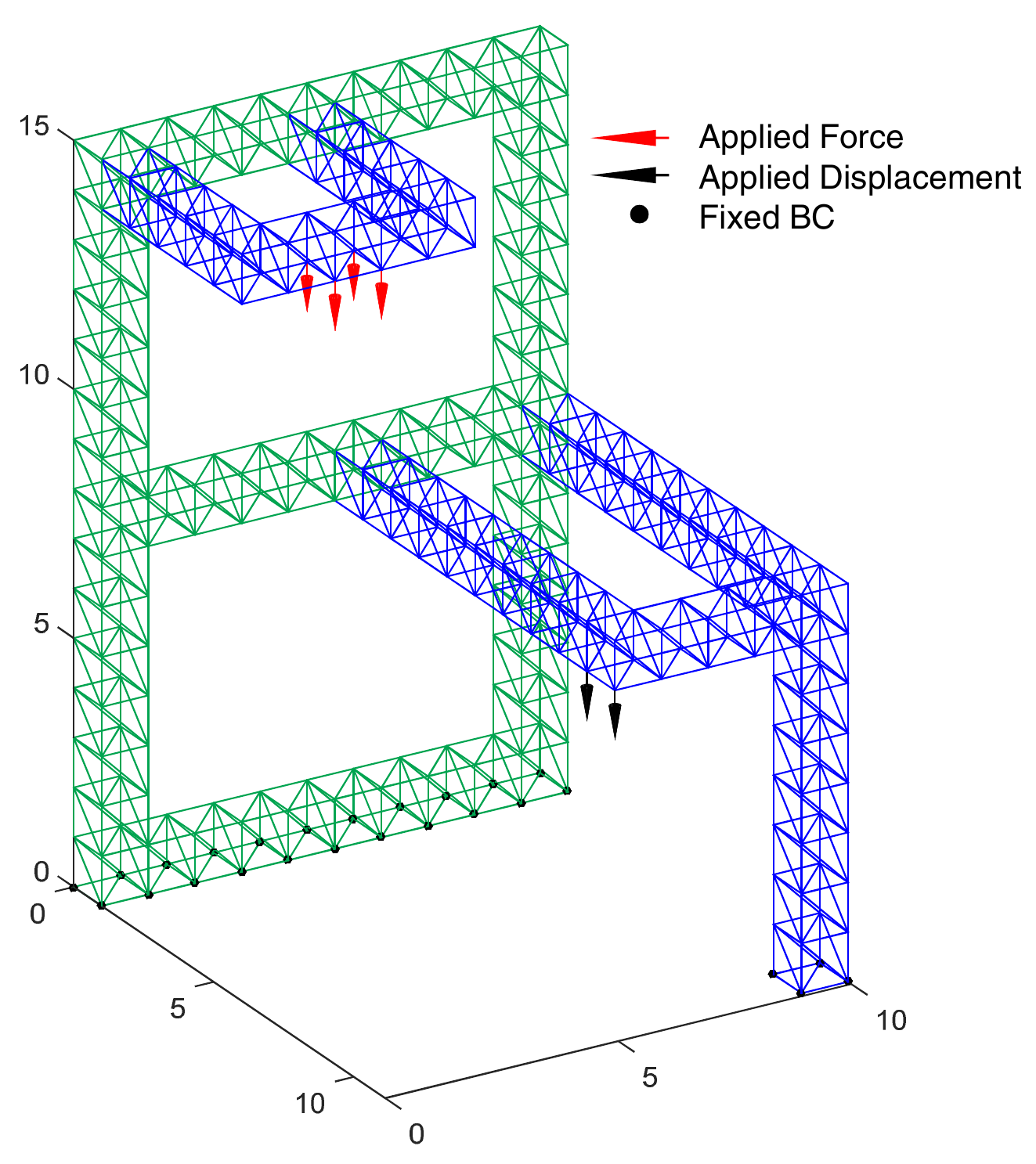}
    \caption{}
\end{subfigure}
\begin{subfigure}[]{0.56\textwidth}
    \centering
    \includegraphics[width=\textwidth]{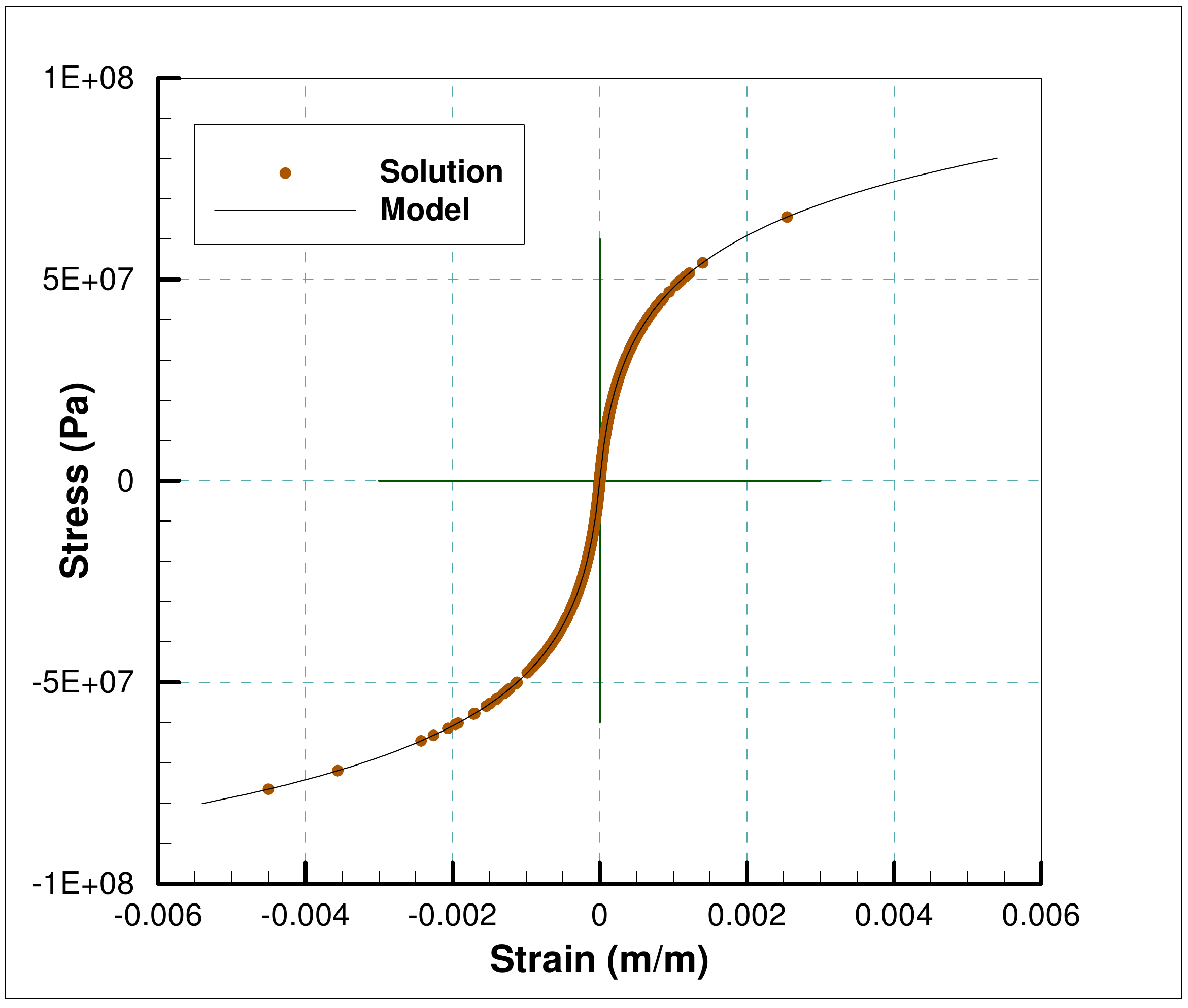}
    \caption{}
\end{subfigure}
    \caption{a) Geometry and boundary conditions of truss test case. b) Base material model with reference solution stress-strain points superimposed.}
    \label{Truss_Model}
\end{figure}


In calculations we consider the specific test case shown in Fig.~\ref{Truss_Model}a. The truss contains 1,246 members and is supported and loaded as shown in the figure. By way of reference, we consider the nonlinear stress-strain relation shown in \ref{Truss_Model}b. A Newton-Raphson solution based on that model is readily obtained. The resulting states of all the members of the truss are shown in Fig.~\ref{Truss_Model}a superimposed on the stress-strain curve in order to visualize the coverage of phase space entailed by the reference solution.

\subsection{Annealing Schedule} \label{AnnealingDefinition}

In calculations, we adopt the annealing schedule formulated in Section~\ref{9iechI}. We start the iteration from from the initial conditions
\begin{equation}
    \frac{1}{\beta^{(0)}}
    =
    \frac{1}{n}
    \sum_{i=1}^n |\bar{z}^{(0)} - z_i|^2 ,
\end{equation}
with
\begin{equation}
    \bar{z}^{(0)}
    =
    \frac{1}{n}
    \sum_{i=1}^n z_i ,
\end{equation}
and
\begin{equation}
    z^{(0)} = \bar{z}^{(0)} .
\end{equation}
As a further control on the annealing rate we set
\begin{equation} \label{LamdaSchedule}
    \beta^{(k+1)}
    =
    \lambda \tilde{\beta}^{(k+1)} +(1-\lambda)\beta^{(k)} ,
\end{equation}
where $\tilde{\beta}^{(k+1)}$ is the result of the recurrence relation (\ref{Vo9Xoa}) and $\lambda$ is an adjustable factor. Alternative strategies for starting and accelerating simulated-annealing iterations are briefly noted in Section~\ref{sec:Improvements}, but a detailed investigation of such alternatives is beyond the scope of this paper (cf., e.~g., \cite{RN125} for a general discussion of simulated-annealing strategies). The iterative solver operates in two distinct phases. The first phase executes the annealing schedule until the values for $\beta$ become large. Subsequent to this initial phase, the algorithm proceeds by distance minimization, as in \cite{RN48}, until convergence is achieved.

\begin{figure}[H]
\begin{subfigure}[b]{0.49\textwidth}
    \centering
    \includegraphics[width=\textwidth]{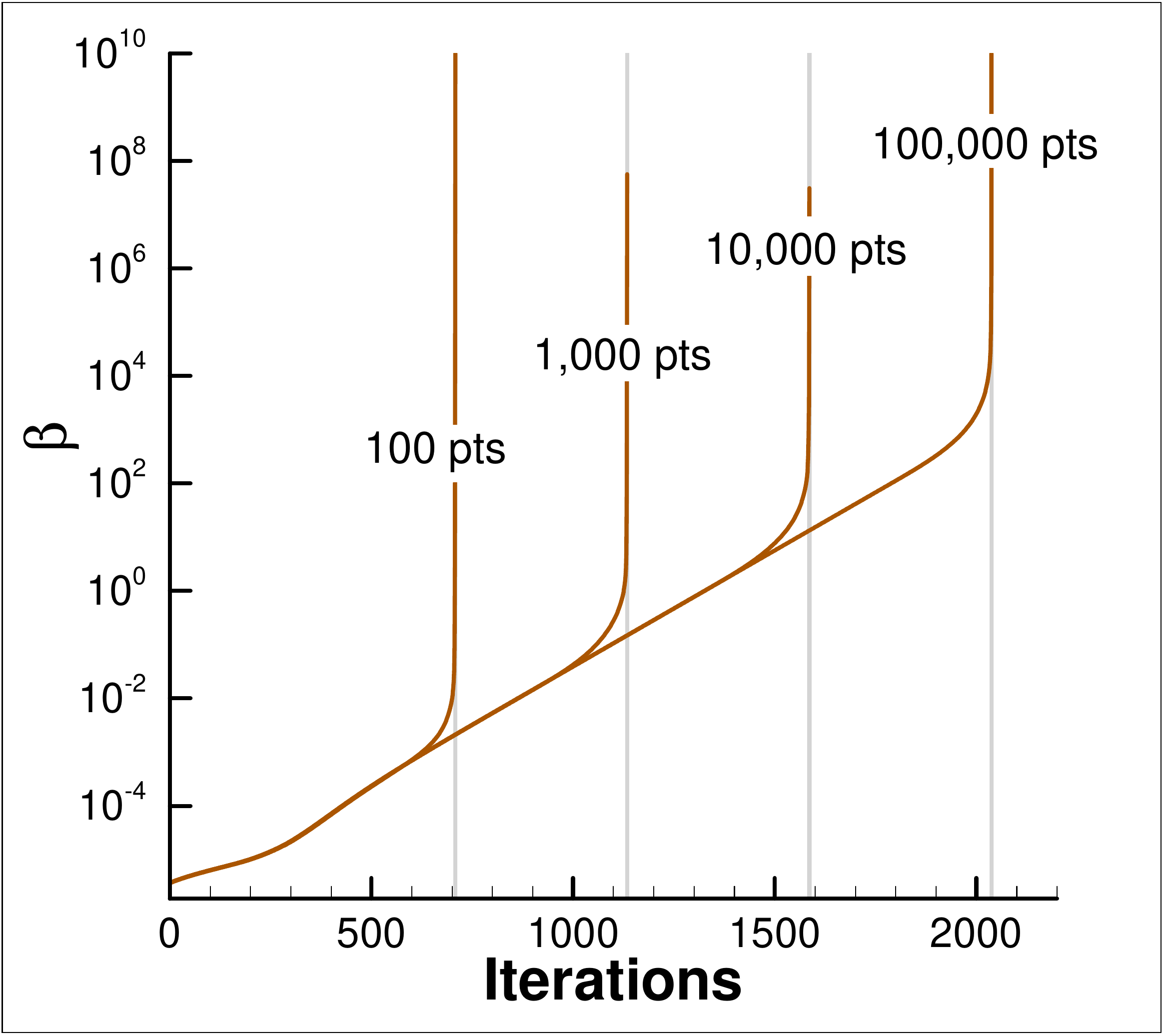}
    \caption{}
\end{subfigure}
\begin{subfigure}[b]{0.49\textwidth}
    \centering
    \includegraphics[width=\textwidth]{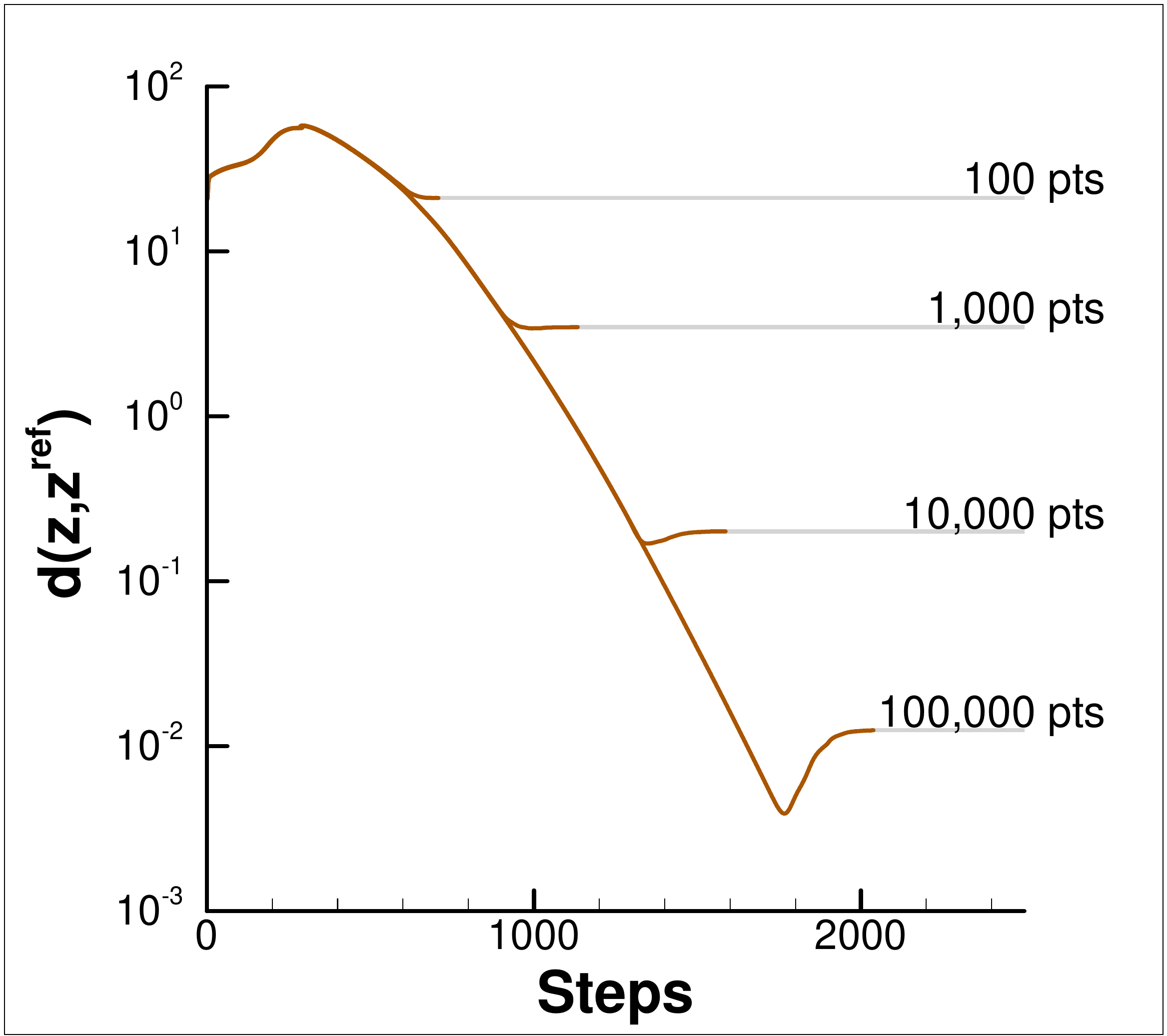}
    \caption{}
\end{subfigure}
    \caption{Truss test case, , $\lambda = 0.01$. a) Evolution of $\beta$ through the annealing schedule for different data set sizes. b) Convergence of the max-ent Data Driven solution to the reference solution for the base model depicted in Fig.~\ref{Truss_Model}b.}
    \label{NoiselessLambda_Beta}
\end{figure}


The performance of the solver just defined is shown in Fig.~\ref{NoiselessLambda_Beta}. In the present test, data sets are generated by spacing the data points evenly over the strain axis and then evaluating the corresponding stress values from the base model depicted in Fig.~\ref{Truss_Model}b. Fig.~\ref{NoiselessLambda_Beta}a shows the evolution of $\beta$ through the annealing schedule for $\lambda = 0.01$. As may be seen from the figure, $\beta$ grows roughly linearly up to a certain, data set size dependent, number of iterations at which point it diverges rapidly. The stepwise convergence of the simulated annealing iteration is shown in Fig.~\ref{NoiselessLambda_Beta}b. As the data set grows in size, the number of iterations to convergence grows correspondingly, as the iteration has to explore a larger data set.

\begin{figure}[H]
\begin{subfigure}[b]{0.49\textwidth}
    \centering
    \includegraphics[width=\textwidth]{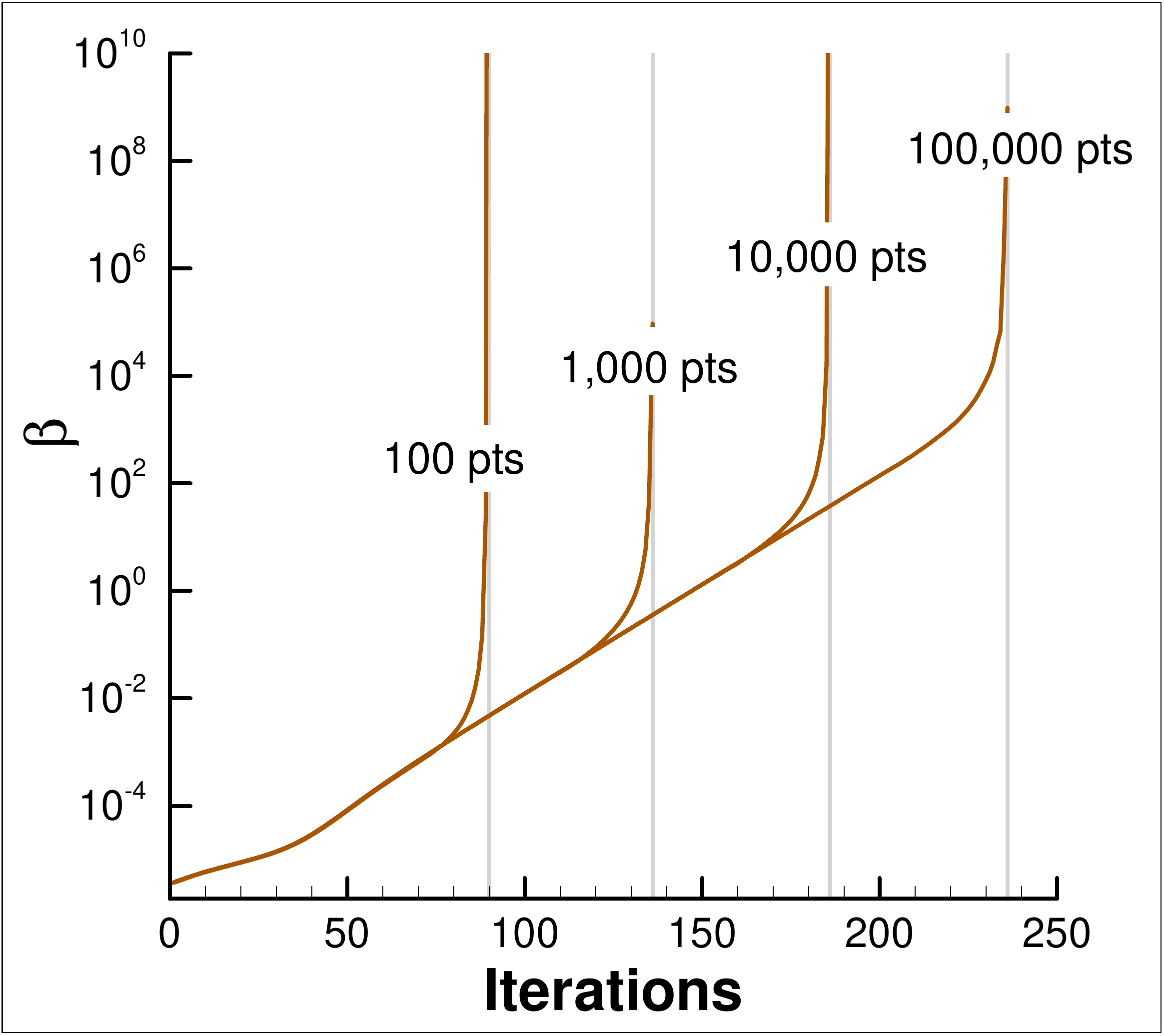}
    \caption{}
\end{subfigure}
\begin{subfigure}[b]{0.49\textwidth}
    \centering
    \includegraphics[width=\textwidth]{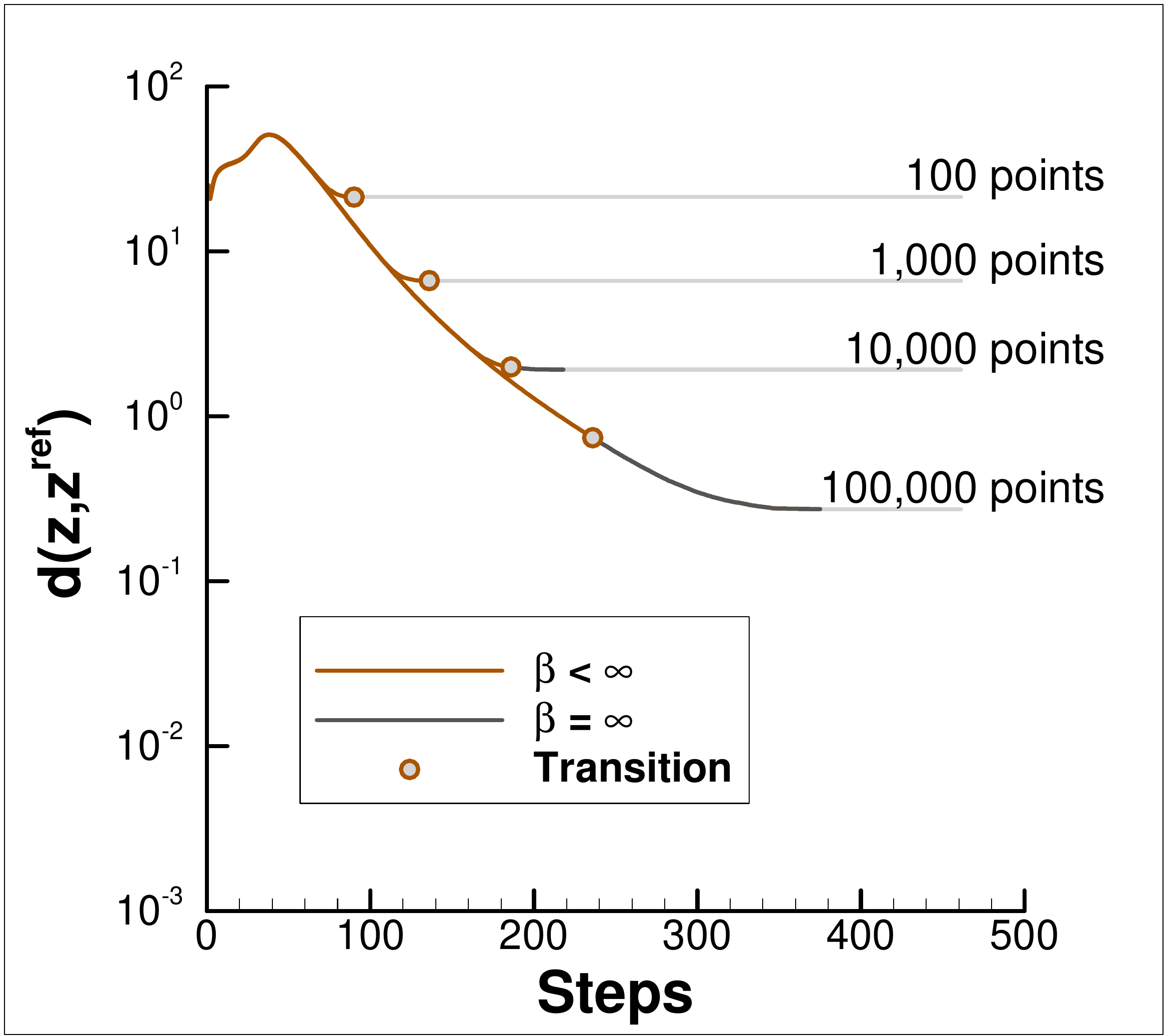}
    \caption{}
\end{subfigure}
    \caption{Truss test case, $\lambda = 0.1$. a) Evolution of $\beta$ through the annealing schedule for different data set sizes. b) Convergence of the max-ent Data Driven solution to the reference solution for the base model depicted in Fig.~\ref{Truss_Model}b.}
    \label{NoiselessLambda_Beta2}
\end{figure}


The influence of the parameter $\lambda$ on the annealing schedule and the solution is illustrated in Fig.~\ref{NoiselessLambda_Beta2}, which corresponds to $\lambda = 0.1$. In general, a larger value of $\lambda$ represents a more aggressive, or faster, annealing schedule, whereas a smaller value represents a more conservative, or slower, annealing schedule. A comparison between Figs.~\ref{NoiselessLambda_Beta} and \ref{NoiselessLambda_Beta2} reveals that, whereas an aggressive annealing schedule indeed speeds up the convergence of the simulated-annealing iteration, it may prematurely freeze the solution around a non-optimal data set cluster, with an attendant loss of accuracy of the solution. Contrariwise, whereas a conservative annealing schedule slows down the convergence of the simulated-annealing iteration, it provides for a more thorough exploration of the data set, resulting in a solution of increased accuracy.

\begin{figure}[H]
\begin{subfigure}[b]{0.49\textwidth}
    \centering
    \includegraphics[width=\textwidth]{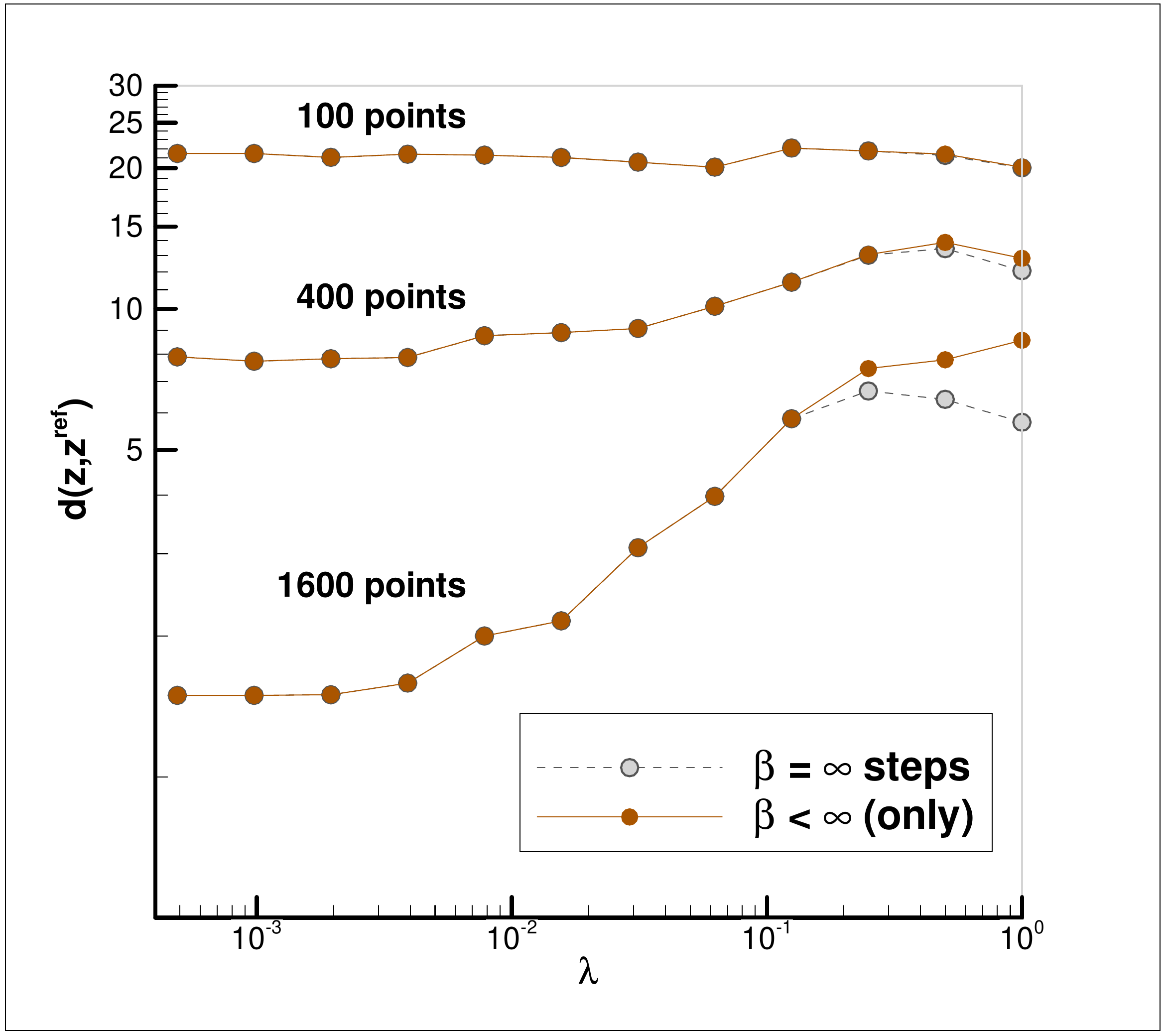}
    \caption{}
\end{subfigure}
\begin{subfigure}[b]{0.49\textwidth}
    \centering
    \includegraphics[width=\textwidth]{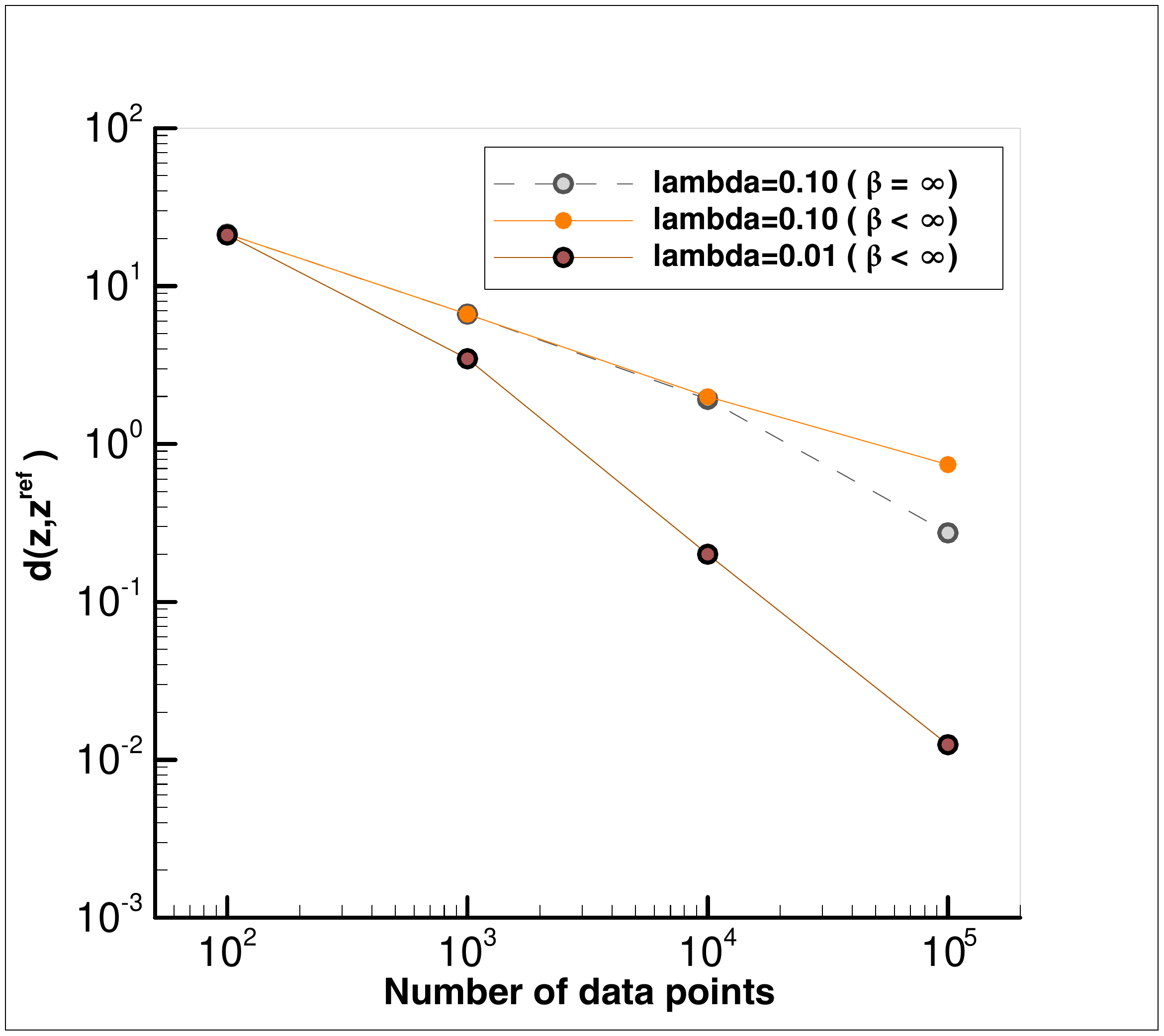}
    \caption{}
\end{subfigure}
    \caption{Truss test case. a) Error in the data driven solution relative to the reference solution as a function of $\lambda$ and data set size. b) Convergence to the reference solution with increasing data set size.}
    \label{Anneal_Effect}
\end{figure}


Further evidence of this annealing speed {\sl vs}.~accuracy trade-off is collected in Fig.~\ref{Anneal_Effect}. Thus, Fig.~\ref{Anneal_Effect}a shows the error in the data driven solution relative to the reference solution as a function of $\lambda$ and data set size. As may be seen from the figure, the data driven solution is relatively insensitive to $\lambda$ for small, or coarse, data sets. This lack of sensitivity owes to the fact that, by virtue of the coarseness of the data set, the simulated-annealing iteration leads to identical, or nearly identical, local data cluster regardless of the value of $\lambda$. By contrast, the range of possible limiting local data clusters increases with the size of the data set. Under these conditions, a conservative annealing schedule is more effective at identifying an optimal, or nearly-optimal, local data set cluster, at an attendant improvement in the accuracy of the solution. Fig.~\ref{Anneal_Effect}a also illustrates the beneficial effect of performing a distance-minimizing iteration after quenching (grey symbols) {\sl vs}.~stopping the iteration upon quenching (orange symbols). Fig.~\ref{Anneal_Effect}b shows the rates of convergence achieved as a function of $\lambda$ and the size of the data set. A theorem presented in \cite{RN48} shows that, for the data sets under consideration, the rate of convergence of distance-minimizing Data Driven solutions with respect to data set size is linear. Fig.~\ref{Anneal_Effect}b suggests that the same rate of convergence is achieved asymptotically by the max-ent Data Driven solutions for sufficiently small $\lambda$.

\subsection{Uniform convergence of a noisy data set towards a classical material model}

\begin{figure}[H]
\begin{subfigure}[]{0.51\textwidth}
    \centering
    \includegraphics[width=\textwidth]{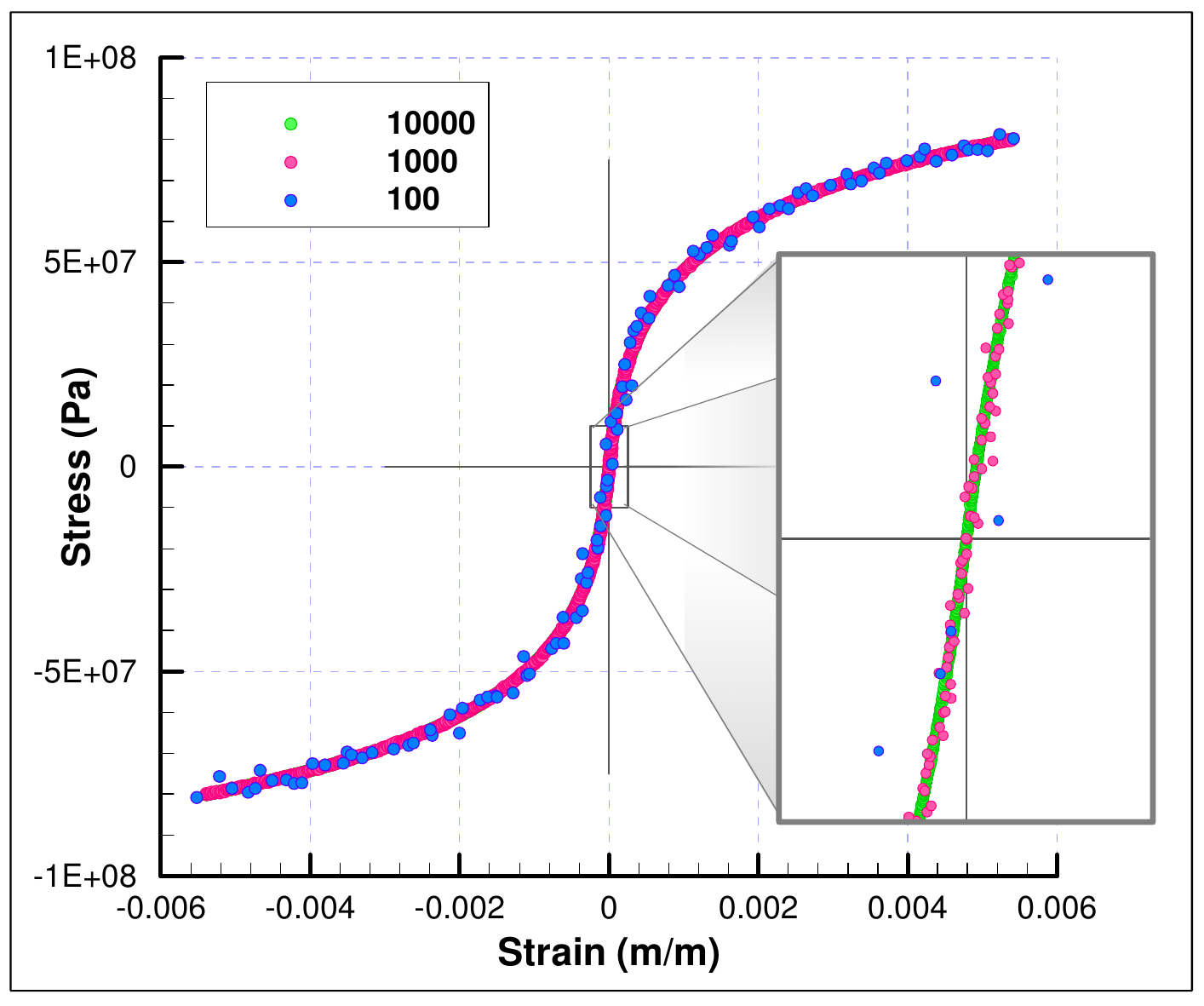}
    \caption{}
\end{subfigure}
\begin{subfigure}[]{0.47\textwidth}
    \centering
    \includegraphics[width=\textwidth]{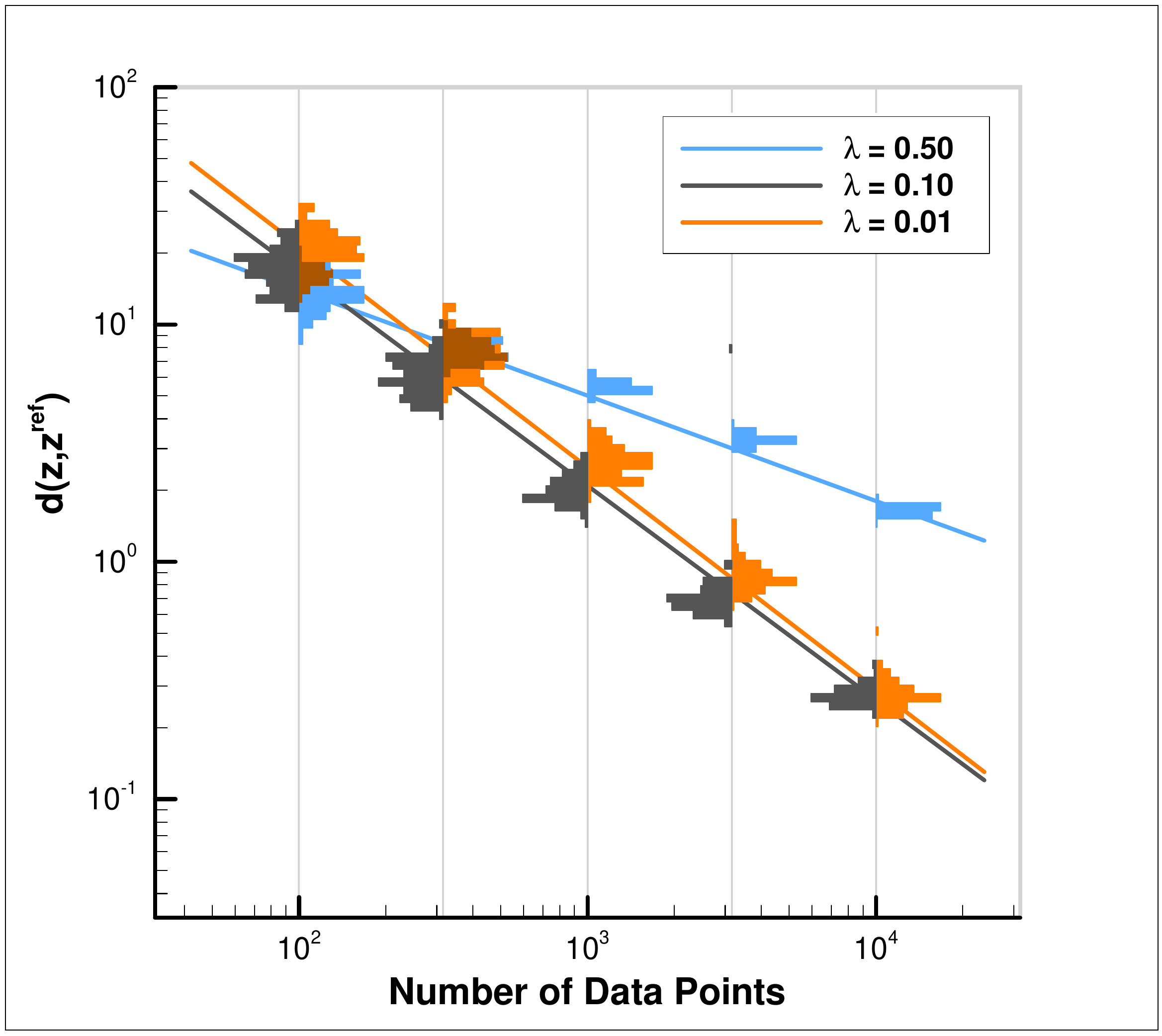}
    \caption{}
\end{subfigure}
    \caption{Truss test case. a) Random data sets generated according to capped normal distribution centered on the material curve of Fig.~\ref{Truss_Model}b with standard deviation in inverse proportion to the square root of the data set size. b) Convergence with respect to data set size of error histograms generated from 100 material set samples.}
    \label{Uniform_Conv}
\end{figure}

Next we consider data sets that, while uniformly convergent to a material curve in phase space, include noise in inverse proportion to the square root of the data set size. To construct a data set consistent with this aim, points are first generated directly from the material curve so that the metric distance between the points is constant. This first sample then has noise added independently pointwise according to a capped normal distribution in both the strain and stress axes with zero mean and standard deviation in inverse proportion to the square root of the data set size. The resulting data sets converge uniformly to the limiting material curve with increasing number of data points. Fig.~\ref{Uniform_Conv}a illustrates the data sets thus generated when the limiting model is as shown in Fig.~\ref{Truss_Model}b.

A convergence plot of error {\sl vs}.~data set size in shown in Fig.~\ref{Uniform_Conv}b, with error defined as the distance between the max-ent Data Driven solution and the classical solution. For every data set size, the plot depicts histograms of error compiled from 100 randomly generated data set samples. We again recall that, given the capped structure of the data sets under consideration, distance-minimizing Data Driven solutions converge to the limiting classical solution as $N^{-1/2}$, with $N$ the size of the data set \cite{RN48}. Surprisingly, an analysis of Fig.~\ref{Uniform_Conv}b suggests that, for sufficiently small $\lambda$, the max-ent Data Driven solutions converge as $N^{-1}$ instead, i.~e., they exhibit a linear rate of convergence with the data set size.

The random sampling of the data sets also raises questions of convergence in probability. It is interesting to note from Fig.~\ref{Uniform_Conv}b that both the mean error and the standard deviation of the error distribution converge to zero with increasing data set size. As already noted, the mean error exhibits a linear rate of convergence. The roughly constant width of the error histograms in log-log coordinates, suggests that the standard deviation of the error also converges to zero linearly with increasing data set size. These two observations together suggest that the error distribution obtained from a capped normal sampling of a material reference curve converges with sample size to the Dirac distribution centered at zero in both mean and in mean square, hence in probability \cite{RN126}.

\subsection{Random data sets with fixed distribution about a classical material model}

A different convergence scenario arises in connection with random material behavior described by a {\sl fixed} probability measure $\mu$ in phase space.  Specifically, given a set $E$ in phase space, $\mu(E)$ is the probability that a fair test return a state $z \in E$. By virtue of the randomness of the material behavior, the solution becomes itself a random variable. We recall that the constraint set $C$ is the set of states $z$ in phase space that are compatible and in equilibrium. When the material behavior is random and is characterized by a probability measure $\mu$ in phase space, the solution must be understood in probabilistic terms and may be identified with the conditional probability $\mu \LL C$ of $\mu$ conditioned to $C$. The corresponding question of convergence then concerns whether the distribution of Data Driven solutions obtained by sampling $\mu$ by means of data sets of increasing size converges in probability to $\mu \LL C$.

\begin{figure}[H]
\begin{subfigure}[]{0.47\textwidth}
  \centering
  \includegraphics[width=\textwidth]{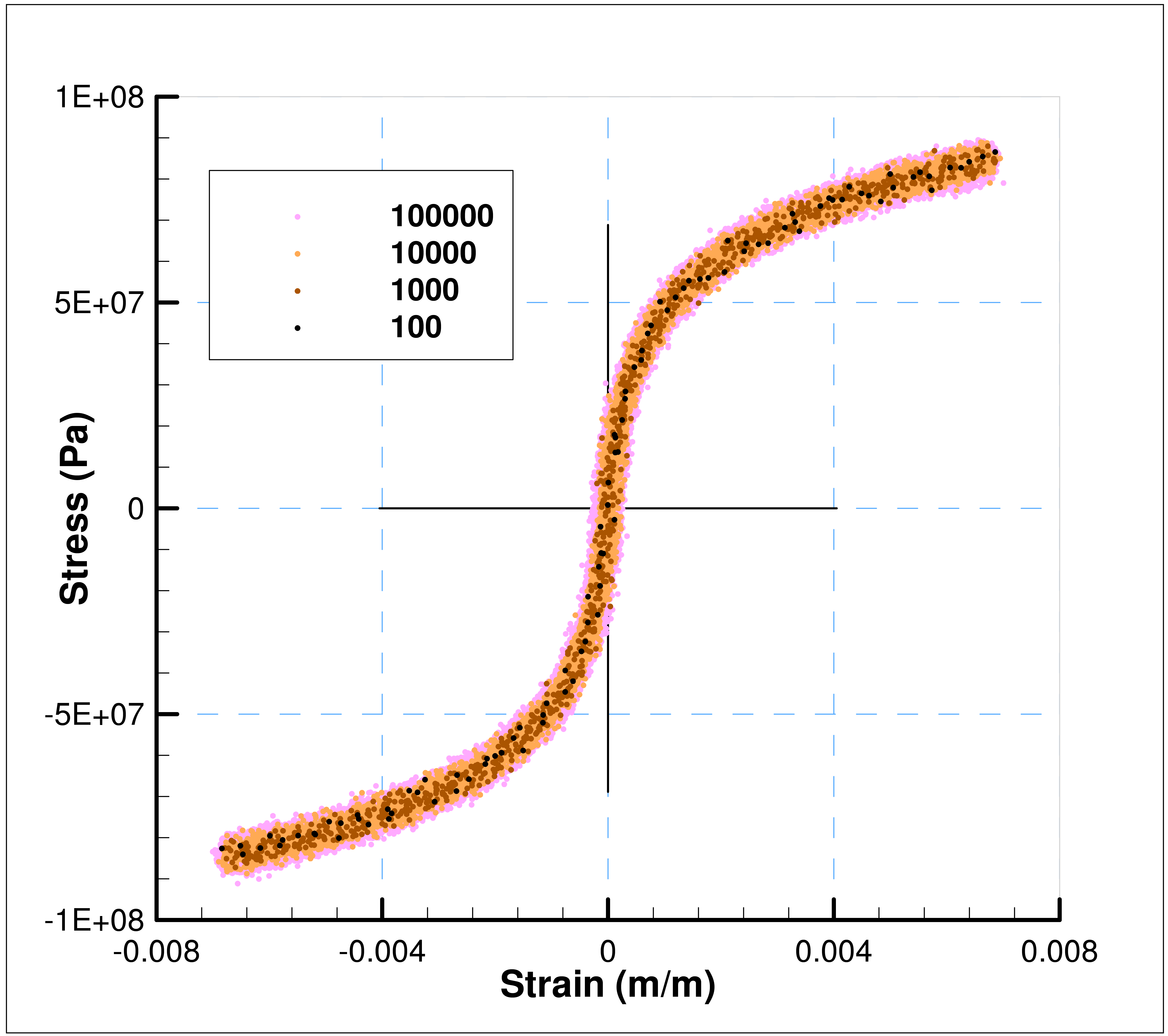}
    \caption{}
\end{subfigure}
\begin{subfigure}[]{0.47\textwidth}
  \centering
  \includegraphics[width=\textwidth]{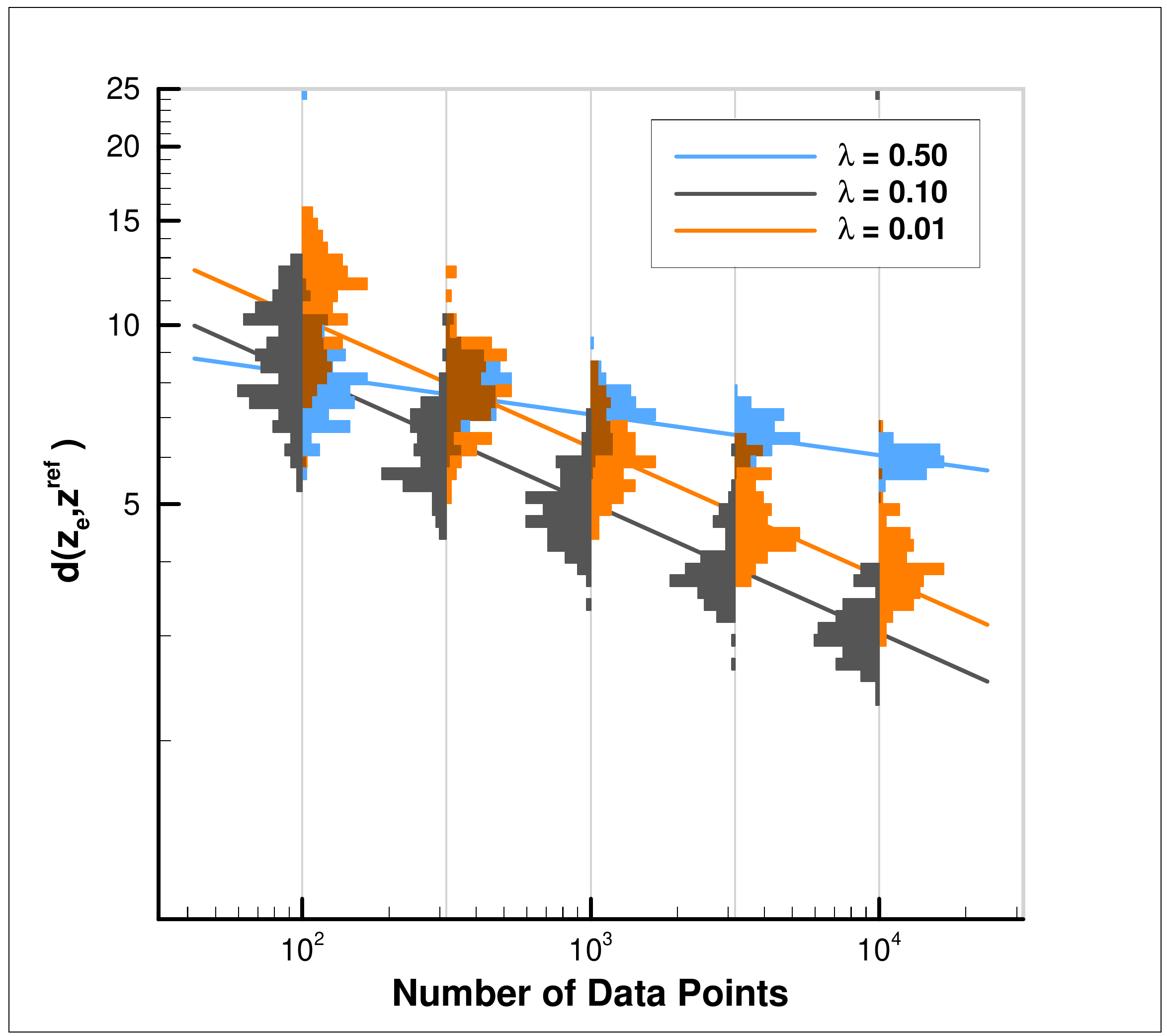}
    \caption{}
\end{subfigure}
  \caption{Truss test case. a) Random data sets generated according to normal distribution centered on the material curve of Fig.~\ref{Truss_Model}b with constant standard deviation independent of the data set size. b) Convergence with respect to data set size of error histograms generated from 100 material set samples.}
  \label{NonUniform_Anneal}
\end{figure}


While a rigorous treatment of convergence in probability is beyond the scope of this paper, we may nevertheless derive useful insights from numerical tests. We specifically assume that $\mu$ is the cartesian product of member-wise measures $\mu_e$ characterizing the material behavior of each bar $e$. Specifically, given a set $E_e$ in the phase space of member $e$, $\mu_e(E_e)$ is the probability that a fair test of member $e$ return a state $z_e \in E_e$. In accordance with this representation, in calculations we generate data sets member-wise from a zero-mean normal distribution that is no longer capped and whose standard deviation is held constant. Fig.~\ref{NonUniform_Anneal}a illustrates the data sets thus generated when the base model is as shown in Fig.~\ref{Truss_Model}b.

Since the probability measure $\mu_e$ is generated by {\sl adding} zero-mean normal random displacements to the base model in phase space, and since the constraint set $C$ is {\sl linear}, the conditional probability $\mu \LL C$ is itself centered on the base model. Hence, its mean value $\bar{z}$ necessarily coincides with the classical solution. This property is illustrated in Fig.~\ref{NonUniform_Anneal}b, which shows a convergence plot of error {\sl vs}.~data set size, with error defined as the distance between the max-ent Data Driven solution and the classical solution. For every data set size, the plot depicts histograms of error compiled from 100 randomly generated data set samples. As may be seen from the figure, the mean value of the histograms converges to zero with data set size, which is indicative of convergence in mean of the sampled max-ent Data Driven solutions. The rate of convergence of the mean error is computed to be of the order of $0.22$. Interestingly, this rate of convergence is considerably smaller than the linear convergence rate achieved for the capped normal noise distributions considered in the preceding section. The slower rate of convergence may be attributable to the wider spread of the data about its mean, though the precise trade-off between convergence and uncertainty remains to be elucidated rigorously. Finally, we note from Fig.~\ref{NonUniform_Anneal}b that, as in the case of capped normal noise, an overly fast annealing schedule results in a degradation of the convergence rate.

\section{Summary and discussion}
\label{1iaFRl}

We have formulated a Data Driven Computing paradigm, which we have termed max-ent Data Driven Computing, that generalizes distance-minimizing Data Driven Computing of the type proposed in \cite{RN48} and is robust with respect to outliers. Robustness is achieved by means of clustering analysis. Specifically, we assign data points a variable relevance depending on distance to the solution and through maximum-entropy estimation. The resulting problem consists of the minimization of a suitably-defined free energy over phase space subject to compatibility and equilibrium constraints. The problem is non-standard in the sense that the relevant Data-Driven free energy is defined jointly over driving forces and fluxes. The distance-minimizing Data Driven schemes \cite{RN48} are recovered in the limit of zero temperature. We have also developed a simulated annealing solver that delivers the solution through a suitably-defined quenching schedule. Finally, we have presented selected numerical tests that establish the good convergence properties of the max-ent Data Driven solutions and solvers.

We conclude by framing Data Driven Computing within the context of past and present efforts to automate the connection between data and material models and expounding on how Data Driven Computing differs from said efforts. We also point out a number of possible enhancements of the approach that define worthwhile directions of further research.

\subsection{Irreducibility to classical material laws}

As already noted, the Data-Driven free energy (\ref{tpG2dU}) and the associated problem (\ref{SoesI5}) are non-standard. Thus, if $z = (\epsilon, \sigma)$, with $\epsilon$ the collection of local states of the system and $\sigma$ the corresponding fluxes, the classical free energy is a function of state of the form $A(\epsilon,\beta)$, i.~e., it is a function of the driving forces and temperature, and the fluxes follow as $\sigma = \partial_\epsilon A(\epsilon,\beta)$. The corresponding classical problem consists of minimizing $A(\epsilon,\beta)$ with respect to the driving forces $\epsilon$ subject to compatibility constraints. Correspondingly, the classical Gibbs energy is a function of state of the form $G(\sigma,\beta)$, i.~e., a function of the fluxes and temperature, and the driving forces follow as $\epsilon = \partial_\sigma G(\sigma,\beta)$. The corresponding classical problem consists of minimizing $G(\sigma,\beta)$ with respect to the fluxes $\sigma$ subject to equilibrium constraints. By contrast, here the relevant free energy (\ref{tpG2dU}) is a function $F(z,\beta)$ defined over the entire phase space, i.~e., is a joint function of the driving forces and fluxes. The corresponding Data-Driven problem consists of minimizing $F(z,\beta)$ with respect to $z$ subject to compatibility and equilibrium constraints simultaneously. Of course, it is possible to define an effective free energy through a partial minimization of $F(z,\beta)$ with respect to the fluxes, i.~e.,
\begin{equation}\label{0Iujla}
    A(\epsilon,\beta)
    =
    \min \,
    \{ F\big((\epsilon, \sigma), \beta\big), (\epsilon, \sigma) \in C \} ,
\end{equation}
with $A = +\infty$ is no minimizer exists. The corresponding effective classical problem then consists of minimizing $A(\epsilon,\beta)$ with respect to the driving forces $\epsilon$. Likewise, it is possible to define an effective Gibbs energy through a partial minimization of $F(z,\beta)$ with respect to the driving forces, i.~e.,
\begin{equation}\label{n9eHlu}
    G(\sigma,\beta)
    =
    \min \,
    \{ F\big((\epsilon, \sigma), \beta\big), (\epsilon, \sigma) \in C \} ,
\end{equation}
with $G = +\infty$ is no minimizer exists. The corresponding effective classical problem then consists of minimizing $G(\sigma,\beta)$ with respect to the fluxes $\sigma$. However, we note that these effective energies are {\sl global} and do not correspond to a classical {\sl local} material law in general. For instance, consider a finite-dimensional problem, such as the truss example developed in the foregoing, with compatibility and equilibrium constraints that can be satisfied identically through the representations
\begin{subequations}
\begin{align}
    &
    \epsilon = B u,
    \\ &
    \sigma = A \varphi ,
\end{align}
\end{subequations}
where $u$ is a displacement vector, $B$ is a discrete strain operator, $\varphi$ is an Airy potential vector and $A$ is a discrete Airy operator, with the  properties
\begin{subequations}
\begin{align}
    &
    B^T A = 0,
    \\ &
    A^T B = 0 ,
\end{align}
\end{subequations}
which identify $B^T$ and $A^T$ as the discrete equilibrium and compatibility operators, respectively. In this representation, the Data-Driven problem becomes
\begin{equation}
    F(Bu,A\varphi,\beta) \to \min!
\end{equation}
The corresponding Euler-Lagrange equations are
\begin{subequations}
\begin{align}
    & \label{R0ucrO}
    A^T \frac{\partial F}{\partial\sigma}(Bu,A\varphi,\beta) = 0,
    \\ & \label{roEs5l}
    B^T \frac{\partial F}{\partial\epsilon}(Bu,A\varphi,\beta) = 0,
\end{align}
\end{subequations}
which represent the discrete compatibility and equilibrium equations, respectively. Evidently, it is now possible to eliminate the Airy potential vector $\varphi$ using the compatibility equations (\ref{R0ucrO}) to define a reduced equilibrium problem in the displacement vector $u$, or, alternatively, eliminate the displacement vector $u$ using the equilibrium equations (\ref{roEs5l}) to define a reduced compatibility problem in the Airy potential vector $\varphi$. However, these reduced problems remain non-classical in that they are {\sl non-local}, i.~e., they do not correspond to any local member-wise material law in general.

\subsection{Material Informatics}

There has been extensive previous work focusing on the application of Data Science and Analytics to material data sets. The field of Material Informatics (cf., e.~g., \cite{RN211, RN210, RN206, RN205, RN203, RN202, RN200, RN199, RN226, RN227}) uses data searching and sorting techniques to survey large material data sets. It also uses machine-learning regression \cite{RN197, RN212} and other techniques to identify patterns and correlations in the data for purposes of combinatorial materials design and selection. These approaches represent an application of standard sorting and statistical methods to material data sets. While efficient at looking up and sifting through large data sets, it is questionable that any real epistemic knowledge is generated by these methods. What is missing in Material Informatics is an explicit acknowledgement of the field equations of physics and their role in constraining and shaping material behavior. By way of contrast, such field equations play a prominent role in the Data Driven Computing paradigm developed in the present work.

\subsection{Material identification}.

There has also been extensive previous work concerned with the use of empirical data for parameter identification in prespecified material models, or for automating the calibration of the models. For instance, the Error-in-Constitutive-Equations (ECE) method is an inverse method for the identification of material parameters such as the Young’s modulus of an elastic material \cite{RN229, RN230, RN231, RN232, RN238, RN241, RN242, RN243, RN244, RN245}. While such approaches are efficient and reliable for their intended application, namely, the identification of material parameters, they differ from Data Driven Computing in that, while material identification schemes aim to determine the parameters of a prespecified material law from experimental data, Data Driven Computing dispenses with material models altogether and uses fundamental material data directly in the formulation of initial-boundary-value problems and attendant calculations thereof.

\subsection{Data repositories}

A number of repositories are presently in existence aimed at data-basing and disseminating material property data, e. g., \cite{NoMaD, MP, KIM, NIST}. However, it is important to note that the existing material data repositories archive parametric data that are specific to prespecified material models. For instance, a number of repositories rely on parameterizations of standard interatomic potentials, such as the embedded-atom method (EAM), and archive data for a wide range of materials systems. Evidently, such data are strongly biased by---and specific to---the assumption of a specific form of the interatomic potential. By way of contrast, Data Driven Computing is based on fundamental, or {\sl model-free}, material data only. Thus, suppose that the problem of interest is linear elasticity. In this case, the field equations are the strain-displacement and the equilibrium relations, and the local states are described by a strain tensor and a corresponding stress tensor. It thus follows that, in this case, model-free fundamental data consists of points in strain-stress space, or {\sl phase space}. By relying solely on fundamental data, Data Driven Computing requires no {\sl a priori} assumptions regarding particular forms, and parameterizations thereof, of material models.

\subsection{Implementation Improvements}
\label{sec:Improvements}

This paper has focused on a particular definition of the annealing schedule as a means to implementing the new class of max-ent Data Driven solvers. It remains easily within the bounds of expectation that improvements in the schedule definition could lead to reductions in the number of iterations and improvements in annealing convergence rates. A number of other implementation improvements are equally worthy of examination. At present, sums over entire data sets for each material point were calculated without simplification or truncation. However, early stages of the annealing schedule could easily be performed on subsampled or summarized data sets due to the nonlocal nature of the calculations. Late stages of the annealing schedule could easily truncate sums over the data set through the use of cutoff radii pegged to $\beta^{-1/2}$. These and other considerations are likely to play an important role in the progression towards efficient and scalable implementations of the method.

\subsection{Data coverage, sampling quality, adaptivity}

Data Driven solvers provide, as a by-product, useful information regarding data coverage and sampling quality. Specifically, suppose that $z$ is a Data Driven solution and $z_e$ is the corresponding local state at material point $e$. Then, the distance $d_e(z_e,E_e)$ supplies a measure of how well the local state $z_e$ is represented within the local material data set $E_e$. For any given material data set, a certain spread in the values of $d_e(z_e,E_e)$ may be expected, indicating that certain local states in a solution are better sampled than others. Specifically, local states with no nearby data points result in high values of $d_e(z_e,E_e)$, indicative of poor coverage by the material data set. Thus, the analysis of the local values $d_e(z_e,E_e)$ of the distance function provides a means of improving material data sets adaptively for particular applications. Evidently, the optimal strategy is to target for further sampling the regions of phase space corresponding to the local states with highest values of $d_e(z_e,E_e)$. In particular, local states lying far from the material data set, set targets for further testing. In this manner, the material data set may be adaptively expanded so as to provide the best possible coverage for a particular application.

\subsection{Data quality, error bounds, confidence}

Not all data are created equal, some data are of higher quality than others. In general, it is important to keep careful record of the pedigree, or ancestry, of each data point and to devise metrics for quantifying the level of confidence that can be placed on the data \cite{RN136}. The confidence level in a material data point $z_i$ can be quantified by means of a confidence factor $c_i\in [0,1]$, with $c_i=0$ denoting no confidence and $c_i=1$ denoting full confidence. The weighting of the data points can then be modified to
\begin{subequations}
\begin{align}
    &
    p_i(z,\beta) = \frac{c_i}{Z(z,\beta)} {\rm e}^{-(\beta/2) d^2 (z,z_i)} ,
    \\ &
    Z(z,\beta) = \sum_{i=1}^n c_i {\rm e}^{-\beta d^2 (z,z_i)} ,
\end{align}
\end{subequations}
which effectively factors the confidence factors into the calculations.  In addition, material data obtained through experimental measurements often comes with error bounds attached. The standard error of a measurement of mean $z_i$  is normally identified with its standard deviation $s_i$. In such cases, assuming the distribution of measurements to be Gaussian we obtain the distribution of weights
\begin{subequations}
\begin{align}
    &
    p_i(z,\beta)
    =
    \frac{1}{Z(z,\beta)} {\rm e}^{-1/2(s_i^2 + 1/2\beta)^{-1} d^2 (z,z_i)} ,
    \\ &
    Z(z,\beta)
    =
    \sum_{i=1}^n {\rm e}^{-1/2(s_i^2 + 1/2\beta )^{-1} d^2 (z,z_i)} .
\end{align}
\end{subequations}
Again, this simple device effectively factors the experimental error bounds into the calculations.

\section*{Acknowledgements}

The support of Caltech's Center of Excellence on High-Rate Deformation Physics of Heterogeneous Materials, AFOSR Award FA9550-12-1-0091, is gratefully acknowledged. We gratefully acknowledge helpful discussions with H.~Owhadi and T.~J.~Sullivan.


\bibliographystyle{plain}
\bibliography{biblio}

\end{document}